\documentclass[
superscriptaddress,
 amsmath,amssymb,
 aps,
]{revtex4-2}
\usepackage[english]{babel}
\usepackage{amsmath}
\usepackage{graphicx}
\usepackage{dcolumn}
\usepackage{bm}
\usepackage{hyperref}
\usepackage{cleveref}
\usepackage{tikzit}
\usepackage{subcaption}
\usepackage{caption}
\tikzstyle{new style 0}=[fill=green, draw=black, shape=circle]
\tikzstyle{red pot}=[fill=red, draw=black, shape=circle]

\tikzstyle{dash}=[-, dashed]
\tikzstyle{blue}=[-, draw=blue]
\tikzstyle{new edge style 0}=[<->, dashed]

\begin{document}

\preprint{APS/123-QED}

\title{Dynamics of $Z_N$ domain walls with bias directions}

\author{Yuan-Jie Li}
\email{liyuanjie23@mails.ucas.ac.cn}
\affiliation{%
International Centre for Theoretical Physics Asia-Pacific, University of Chinese Academy
of Sciences, 100190 Beijing, China
}%
\affiliation{%
Taiji Laboratory for Gravitational Wave Universe, University of Chinese Academy of Sciences, 100049 Beijing, China
}%
\author{Jing Liu}
\email{liujing@ucas.ac.cn}
\affiliation{%
International Centre for Theoretical Physics Asia-Pacific, University of Chinese Academy
of Sciences, 100190 Beijing, China
}%
\affiliation{%
Taiji Laboratory for Gravitational Wave Universe, University of Chinese Academy of Sciences, 100049 Beijing, China
}%
\author{Zong-Kuan Guo}
\email{guozk@itp.ac.cn}
\affiliation{%
Institute of Theoretical Physics, Chinese
Academy of Sciences, P.O. Box 2735, Beijing 100190, China
}%
\affiliation{%
School of Physical Sciences, University of Chinese Academy of Sciences, No.19A Yuquan
Road, Beijing 100049, China
}%
\affiliation{%
School of Fundamental Physics and Mathematical Sciences, Hangzhou Institute for Advanced Study, University of Chinese Academy of Sciences, Hangzhou 310024, China
}%


\begin{abstract}
The spontaneous breaking of a discrete symmetry can lead to the formation of domain walls in the early universe. In this work, we explore the impact of bias directions on the dynamics of \(Z_N\) domain walls, mainly focusing on the \(N = 3\) model with a biased potential. Utilizing the Press-Ryden-Spergel method, we numerically investigate the dynamics of domain walls with lattice simulations. We find notable differences in the dynamics of domain walls due to bias directions. Our results indicate that the annihilation time depends not only on the vacuum energy difference \( \delta V \) but also on bias directions described by the relative potential difference~\( \zeta \).

\end{abstract}

\maketitle

\section{introduction}\label{introduction}
The spontaneous symmetry breaking plays an important role in theoretical physics, especially in particle physics and cosmology. When a symmetric system transitions into a lower-energy state that does not preserve the initial symmetry, this process gives rise to topological defects~\cite{kibble1976topology,Vilenkin:2000jqa}. Among these defects, domain walls arise from the spontaneous breaking of a discrete symmetry, such as the $Z_N$ symmetry. 

Domain walls are sheet-like, two-dimensional structures that separate regions (domains) corresponding to different vacua. 
The dynamics of domain walls in the early Universe provide an exciting opportunity to probe physics beyond the Standard Model. As domain walls evolve, their annihilation generates gravitational waves (GWs), imprinting detectable signatures on the stochastic GW background. These signals serve as unique probes of high-energy phenomena and can be tested through next-generation GW detectors, offering constraints on particle physics models that predict the domain wall formation~\cite{Li:2023yzq,Bian:2022qbh,saikawa2017review}. {Previous} studies, such as Hiramatsu \textit{et al.}~\cite{Hiramatsu:2010yz}, show that domain walls are a important cosmological source of GWs. Detecting stochastic GWs produced by domain walls is one of the key scientific objectives of GW detection experiments.

A paradigmatic framework for the discrete symmetry breaking involves $Z_N$ symmetries, which yield $N$ degenerate vacua. Domain walls form at interfaces between these vacua, with their stability and dynamics governed by the underlying potential. Axion models, where the $Z_N$ symmetry emerges naturally, have gained prominence not only as solutions to the strong CP problem but also as viable dark matter candidates~\cite{Sikivie:1982qv,kawasaki_axion_2015,Marsh:2015xka,Gelmini:2022nim,Gelmini:2021yzu}. Scalar field extensions of the standard model, featuring discrete potential minima, further illustrate the interplay between the symmetry breaking and cosmological phenomena such as inflation and dark matter production~\cite{linde1994hybrid}. Additionally, $Z_N$ symmetries also appear in theories of flavor hierarchies~\cite{king2017unified, xing2020flavor, Gelmini:2020bqg} and supersymmetric model building~\cite{ibanez1992discrete}, highlighting their versatility in addressing open questions across energy scales.

Domain walls typically exhibit scaling solutions, where their energy density scales with the background expansion~\cite{Press:1989yh,garagounis2003scaling,oliveira2005cosmological,avelino2005one,leite2011scaling,leite2013accurate,martins2016extending}. This self-similar evolution pattern, extensively verified through numerical lattice simulations~\cite{leite2011scaling,leite2013accurate,martins2016extending}, creates the well-known domain wall problem: prolonged network survival would inevitably dominate the Universe's energy density, conflicting with precision constraints from cosmic microwave background measurements and the large-scale structure~\cite{zel1974cosmological}.
The introduction of vacuum degeneracy-breaking bias terms~\cite{Srivastava:1999br,larsson1997evading} resolves this by inducing pressure asymmetries that drive the domain wall annihilation~\cite{gelmini1989cosmology,vilenkin1981gravitational}.
Such bias terms can arise from physics at the Planck scale~\cite{kamionkowski1992planck,holman1992solutions,dobrescu1997strong,barr1992planck,dine1992problems}.

A main goal of our research is to understand how bias directions influence the dynaimcs and  annihilation time of domain walls. Previous studies~\cite{Hiramatsu:2010yn,Hiramatsu:2012sc,kawasaki_axion_2015,Kawasaki:2013iha,Li:2023gil,Gelmini:2021yzu}, have analyzed the evolution and the GW radiation of $Z_N$ domain walls but have not systematically explored the role of bias directions. This effect is important because bias directions could affect the annihilation time so that domain walls could introduce distinctive features in the GW spectrum, such as a double peak structure, which may serve as a novel observational signature~\cite{wu2022collapsing}.

The simplest $Z_2$ domain walls have been studied in detail (see~\cite{saikawa2017review} for a review), while numerical studies on domain walls beyond $Z_2$ symmetry remain relatively scarce~\cite{hiramatsu2013axion,kawasaki_axion_2015,Hiramatsu:2010yn}. Moreover, in order to investigate the impact of bias directions on the evolution of domain walls, it is required to study domain walls beyond the $Z_2$ symmetry. Therefore, in this paper, we focus on the dynamics of $Z_3$ domain walls as an illustrative example.

While domain walls and their GW signals have been extensively studied, previous simulation studies have not systematically examined the following aspects.
\begin{enumerate}
    \item The influence of bias directions on the annihilation time of domain walls. Wu et al.~\cite{wu2022collapsing} estimate the annihilation time of $Z_3$ domain walls. This estimate is derived from simulation results that focused on the annihilation time of the $Z_2$ domain walls, as reported in~\cite{kawasaki_axion_2015, saikawa2017review}.
    \item The individual evolution of each type of domain wall. Previous studies have typically considered the evolution of the total area of  domain walls rather than the evolution of the area of each type of domain wall.
\end{enumerate}
Our study fills these gaps by performing numerical simulations to explicitly investigate these effects.

This paper is organized as follows. In Sec.~\ref{scaling solution}, we briefly review scaling solutions of domain walls and the estimation of the annihilation time of $Z_2$ domain walls. In Sec.~\ref{sec:the Model}, we describe the model used in our simulations. In Sec.~\ref{set up}, we outline the basic setup of our numerical simulations. Sec.~\ref{sec:Evolution of Domain Walls} presents our simulation results in the form of field configurations and evolution of the area density. In Sec.~\ref{sec:Estimation of the Time of Annihilation}, we provide a semi-analytical estimate of the annihilation time of domain walls, and explain its behavior and the dynamics of domain walls. Finally, in Sec.~\ref{summary}, we summarize our results and discuss its implications for future research. 

\section{Scaling solutions and annihilation of domain walls}\label{scaling solution}

We provide an overview of the general dynamics of domain walls, including their scaling solutions and the process of annihilation.

After the formation of domain walls, their evolution enters a stable phase, where the average number of walls per Hubble volume remains constant throughout the expansion of the Universe. This property, known as the scaling behavior, has been verified both analytically~\cite{Avelino:2005kn,Hindmarsh:1996xv,Hindmarsh:2002bq} and numerically~\cite{Press:1989yh,Garagounis:2002kt,Avelino:2005pe,Oliveira:2004he} for domain walls arising from the spontaneous breaking of $Z_2$ symmetry. For domain walls associated with the symmetries beyond $Z_2$, the scaling behavior has also been tested in~\cite{Li:2023gil,Hiramatsu:2012sc,Hiramatsu:2010yn,Kawasaki:2014sqa,ryden1990evolution}.

In the scaling regime, the energy density of domain walls evolves as
\begin{eqnarray}
    \rho \sim \sigma / t,
\end{eqnarray}
where $\sigma$ is surface energy density, which is a constant over time, and $t$ is cosmic time. 
This is equivalent to
\begin{eqnarray}
     A/V \propto \tau^{-1},
\end{eqnarray}
where \(A/V\) is the comoving area density of domain walls, $\tau$ is the conformal time. We introduce the area parameter~$\mathcal{A}$ to measure the scaling property of domain walls,
\begin{eqnarray}
    \mathcal{A}(\tau) = \frac{A\tau}{V}.
\end{eqnarray}
In the scaling regime, the area parameter remains approximately constant.

Domain walls will eventually annihilate in the cases the potential includes a bias term. Here, we briefly review the analytical estimate of the annihilation time for domain walls arising from the spontaneous $Z_2$ symmetry breaking with a bias term. The pressure acting on domain walls toward the false vacuum is given by  
\begin{eqnarray}
    p_V \sim \delta V,
\end{eqnarray}
where \(\delta V\) is the potential difference across the domain walls. On the other hand, when domain walls evolve to the horizon scale, the tension pressure is  
\begin{eqnarray}
    p_T \sim \frac{\sigma}{t}.
\end{eqnarray}
Since $p_{T}$ decreases with time while $p_{V}$ remains constant, domain walls begin to annihilate when these two pressures become comparable~\cite{Larsson:1996sp}. Thus, we can estimate the typical annihilation time of domain walls as  
\begin{eqnarray}
    t_{\text{ann}} \sim \frac{\sigma}{\delta V}.
\end{eqnarray}

However, when the symmetries of the potentials extend beyond $Z_2$, the estimate of $t_{\text{ann}}$ requires modification. In this case, one vacuum may connect to multiple vacua, and the pressure differences on either side of different types of domain walls can also be different. As a result, the pressure responsible for the annihilation of false vacua no longer originates solely from the true vacuum. As will be demonstrated by the simulation results later, the annihilation times of different types of domain walls also depend on bias directions.

\section{Model}\label{sec:the Model}

{We consider a complex scalar field $\phi$ with a $Z_N$ symmetry ($N > 2$), motivated by axion models originally proposed to solve the strong CP problem~\cite{Peccei:1977hh, Peccei:1977ur, Wilczek:1977pj,dine1981simple,Zhitnitsky:1980tq}. Such models admit multiple degenerate vacua and support domain walls attached to cosmic strings. Beyond QCD axions, general axion-like particles have been extensively studied as dark matter candidates~\cite{Weinberg:1977ma,Abbott:1982af,Dine:1982ah,Preskill:1982cy,Gelmini:2022nim,Gelmini:2021yzu,Gelmini:2023ngs}. Other $Z_N$ symmetric potentials beyond axion models have also been proposed~\cite{wu2022collapsing,Wu:2022tpe}.}

{The Lagrangian is given by
\begin{align}
    \mathcal{L} = \frac{1}{2} \partial_\mu \phi^* \partial^\mu \phi - V(\phi),
\end{align}
where the scalar potential takes the form
\begin{align}\label{potential}
    V(\phi) = \frac{\lambda}{4} (|\phi|^2 - \eta^2)^2 + \frac{m^2 \eta^2}{N^2}\left(1 - \frac{|\phi|}{\eta} \cos(N\theta)\right) + V_{\text{bias}}.
\end{align}
Here, $\lambda$ is the self-coupling constant, $\eta$ denotes the vacuum expectation value of modulus of the scalar field after \( U(1) \) symmetry breaking, $N$ is the domain wall number, and $\theta$ is the phase of $\phi$ ($\phi = |\phi|e^{i\theta}$). The third term, $V_{\text{bias}} = \Xi \eta^3 (\phi e^{-i\delta} + \text{h.c.})$, introduces a small explicit breaking of the $Z_N$ symmetry, with $\Xi$ and $\delta$ controlling the magnitude and direction of the bias, respectively.}

{The first term in Eq.~\eqref{potential} leads to spontaneous $U(1)$ symmetry breaking and the formation of cosmic strings. As the Universe cools, the second term becomes relevant, explicitly breaking $U(1)$ down to $Z_N$ and generating domain walls attached to strings~\cite{Sikivie:1982qv,Vilenkin:2000jqa}. The domain wall tension is given by
\begin{align}
    \sigma = \frac{8m\eta^2}{N^2}.
\end{align}
After the formation of domain walls, their energy density quickly surpasses that of cosmic strings, and the subsequent dynamics of the system become dominated by the domain walls~\cite{Li:2023gil}. Therefore, the energy contribution from the strings is neglected in the following analysis.}

{The bias term destabilizes domain walls, ensuring their eventual annihilation and avoiding the domain wall problem. Since the detailed shape of the potential is not crucial for evolution of domain walls~\cite{Krajewski:2021jje}, we adopt the linear term in Eq.~\eqref{potential} as a representative case.}

{In this study, we focus on $N=3$. For clarity, the three vacua are labeled in ascending order of their potential energies as Vacuum 0, Vacuum 1, and Vacuum 2. The corresponding potential value of Vacuum $i$ is denoted by $V_i$. We define the potential differences $\Delta V_{ij} = |V_i - V_j|$ ($i<j$), and refer to domain walls connecting Vacua $i$ and $j$ as $ij$-domain walls ($ij$-DWs).}

{One approach to investigating the impact of $\delta$ on the annihilation time is to fix the magnitude of the bias $\Xi$ and compare the annihilation times of domain walls for different $\delta$ values. According to previous studies~\cite{Kawasaki:2014sqa, Hiramatsu:2010yn}, the annihilation time of domain walls is related to the potential difference $\delta V$ between the vacua, $t_{\mathrm{ann}} \propto \delta V^{-1}$. However, since $\delta V$ depends on $\delta$, to isolate the effect of \( \delta \) on the annihilation time  from variations in \( \delta V \), we fix $\Delta V_{02}$ constant and leave $V_1$ as a variable.}

\begin{figure}[ht]
    \begin{subfigure}[t]{0.4\textwidth}
        \includegraphics[height=5.5cm]{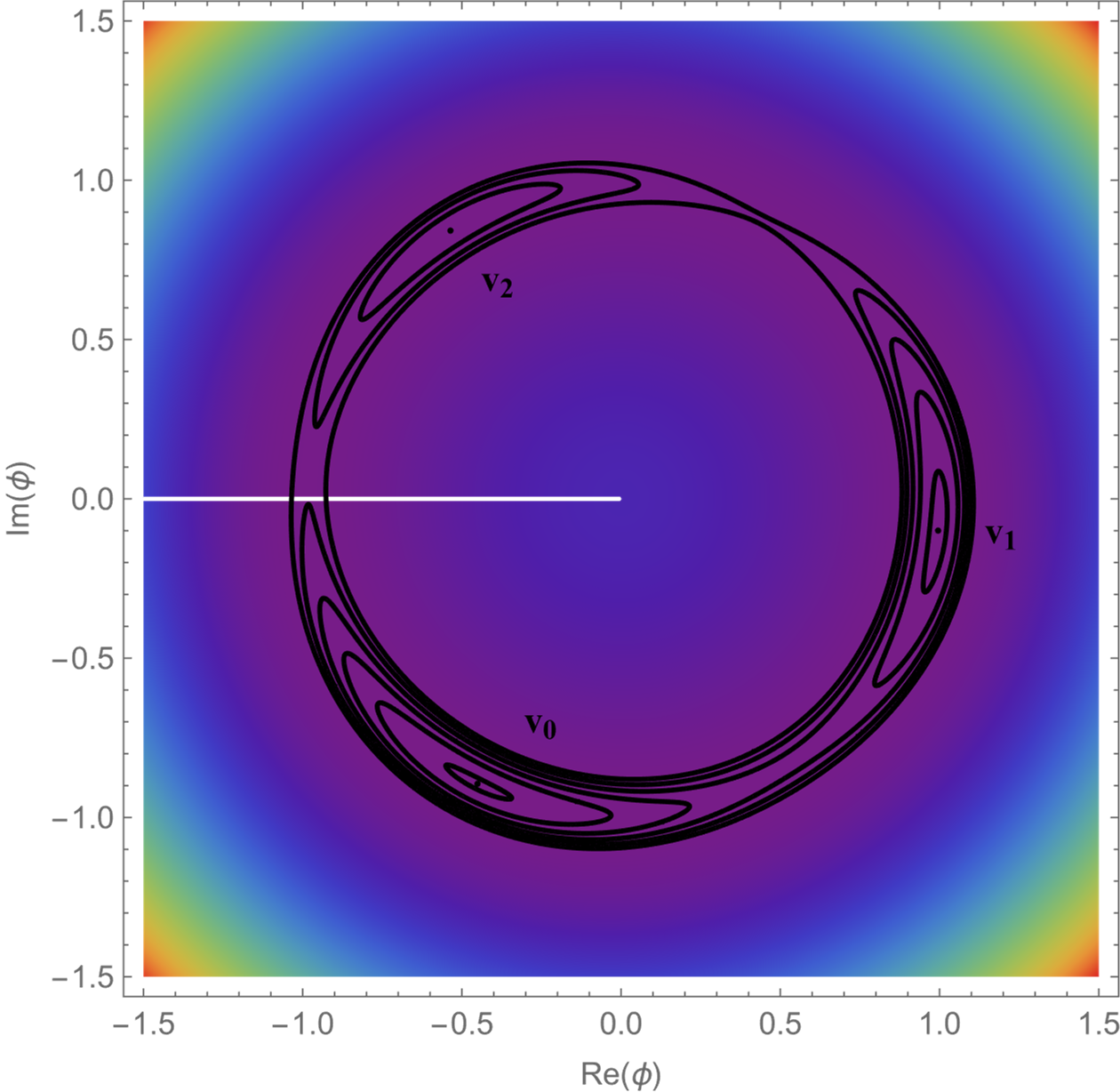}
        \caption{Contour plot of the potential in Eq.~\eqref{potential_pro}, with parameters \(N = 3\), \({m}/(\sqrt{\lambda} \eta) = 0.2\), \(\Xi_1 = -0.0005\), and \(\Xi_2 = 0.004\). The potential increases progressively in the vacua near the phases \(4\pi/3\), 0, and \(2\pi/3\), labeled as \(v_0\), \(v_1\), and \(v_2\), respectively.}
        \label{potential_fig}
    \end{subfigure}
    \hfill
    \begin{subfigure}[t]{0.56\textwidth}
        \includegraphics[height=5cm]{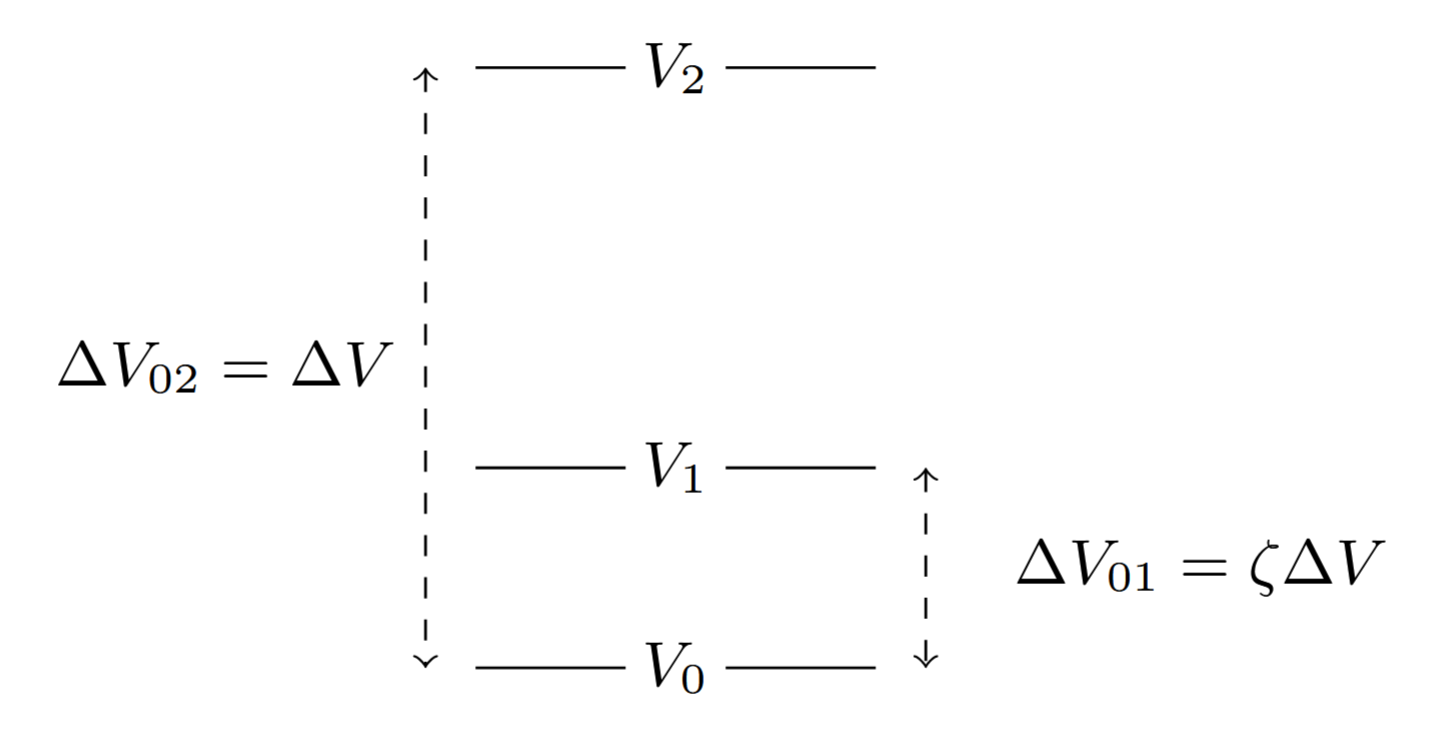} 
        \caption{Diagram of the energy levels for three non-degenerate vacua. \( V_0 \) is the true vacuum, and \( V_1 \), \( V_2 \) are false vacua. \( \Delta V \) denotes the potential difference between \( V_2 \) and \( V_0 \), and \( \zeta = {\Delta V_{01}}/{\Delta V} \) characterizes the relative difference.}
        \label{fig:energy_levels}
    \end{subfigure}
    \caption{(Left) The potential landscape and (right) energy level diagram for the three vacua structure.}
\end{figure}

{We adopt an equivalent parametrization of the bias term,
\begin{eqnarray}\label{potential_1}
    V_{\text{bias}}=  \lambda\eta^3 (\Xi_1 \phi_1 + \Xi_2 \phi_2),
\end{eqnarray}
where $\phi_1 = \mathrm{Re}(\phi)$ and $\phi_2 = \mathrm{Im}(\phi)$. By choosing $\Xi_2>0$ and a small $|\Xi_1|$, the Vacua $0$, $1$ and $2$, correspond to phases near $\theta\sim4\pi/3$, $0$, and $2\pi/3$, respectively, as shown in Fig.~\ref{potential_fig}. Here, the symbol ``$\sim$'' reflects that the vacua are shifted slightly due to the bias term. Varying $\Xi_1$ while keeping $\Xi_2$ fixed allows us to change $V_1$ without changing $\Delta V_{02}$.}

{We set $\Delta V = \Delta V_{02}$ as the largest potential difference and introduce a relative parameter $\zeta \equiv \Delta V_{01}/\Delta V_{02}$ to characterize the potential hierarchy. By varying $\zeta$, we study the impact of bias directions on domain wall evolution. The potential landscape and corresponding energy levels are illustrated in Fig.~\ref{fig:energy_levels}.}

The standard evolution of a scalar field is governed by the Klein-Gordon equation,
\begin{eqnarray}\label{EKG}
    \phi''+2\mathcal{H}\phi'-\nabla_\mathbf{x}^2\phi=-a^2\frac{\partial V}{\partial\phi},
\end{eqnarray}
where a prime denotes the derivative with respect to the conformal time, and $\nabla_\mathbf{x}^2$ is the Laplacian in the comoving coordinates. This equation implies that the physical thickness of domain walls remains constant throughout their evolution, as derived from $d^2\phi/dz^2-dV/d\phi=0$. As the Universe evolves, their comoving width $\delta_c$ decreases as $\delta_c \propto a^{-1}(t)$, where $a(t)$ is the scale factor.

\section{Simulation setup}\label{set up}
Now we briefly summarize the setup of our numerical simulations. For more details, please refer to Appendices~\ref{EoM}, \ref{IC}, \ref{sec:Parameters} and \ref{Cal of AD}.

Lattice simulations are performed in a cubic box with \( \mathrm{N}^3 \) grid points and a comoving volume \( V = L^3 \), where $\mathrm{N}$ is the number of grid points along each dimension, \( L \) denotes the comoving box size. We use \( \mathrm{N} \) in math roman font to distinguish it from the order of symmetry, \( N \), used in \( Z_N \). This box expands with the evolving scale factor~\( a(t) \), simulating the expansion of the Universe. Without loss of generality, we perform lattice simulations in the radiation-dominated era, where \( a \propto \tau \).

Our simulations were performed using the package \textit{CosmoLattice}{(version 1.2)}, employing the second-order Velocity-Verlet algorithm~\cite{figueroa2021art,figueroa2023cosmolattice}.

\subsection{PRS method}\label{PRS Method}

The annihilation times of different domain walls can vary widely depending on bias directions. This variation poses challenges for simulations, as it demands a larger dynamical range to accurately track domain walls behavior and determine their annihilation times. 

In simulations, it is essential to maintain the thickness of domain walls several times the spacing of the grid, expressed as $\delta_c \geq n L / \mathrm{N}$, where $n$ is a positive integer.  However, as domain walls thin in comoving coordinates over time, meeting this requirement becomes increasingly difficult. Moreover, the Hubble horizon must remain smaller than the simulation box size, further constraining the dynamical range. 

For instance, in simulations with $\mathrm{N}=512$, the expansion factor $a_f/a_i$ typically reaches only about a dozen, where $a_i$ and $a_f$ represent the initial and final scale factors,respectively. However, to accurately study annihilation times, the expansion factor must be on the order of hundreds, which is far beyond what standard methods can achieve.

To overcome these limitations, we employ the Press-Ryden-Spergel (PRS) method~\cite{Press:1989yh}. This method maintains a constant comoving domain wall thickness throughout simulations, significantly extending the achievable dynamical range. Additionally, it preserves the tension of domain walls, ensuring that their large-scale dynamics remains consistent with physical expectations. The PRS method has been rigorously validated~\cite{Press:1989yh, Sousa:2010zza} and widely applied in studies of the evolution of topological defects~\cite{Correia:2014kqa, Correia:2018tty, Coulson:1995nv, Leite:2011sc, Krajewski:2021jje}.

In the PRS method, the equation of motion~\eqref{EKG} is modified as follows:
\begin{eqnarray}
    \phi'' + \alpha \mathcal{H} \phi' - \nabla_{\mathbf{x}}^2 \phi = -a^\beta \frac{\partial V}{\partial \phi}.
\end{eqnarray}
To preserve the dynamics of the planar domain walls, $\alpha + \beta / 2$ must equal 3~\cite{Press:1989yh}. To maintain the comoving thickness constant, we set $\beta = 0$, which implies $\alpha = 3$. This simplifies the equation to
\begin{eqnarray}\label{eq:prs}
    \phi'' + 3 \mathcal{H} \phi' - \nabla_{\mathbf{x}}^2 \phi = -\frac{\partial V}{\partial \phi}.
\end{eqnarray}
Our simulations use this modified form, enabling us to achieve the necessary dynamical range while preserving accurate dynamics of domain walls.

\subsection{Redefinition of variables}\label{sec:Program variables}

Our simulation calculations are performed using dimensionless quantities, referred to as program variables. These dimensionless quantities are defined as
\begin{eqnarray}
\tilde{\tau} = \sqrt{\lambda} \eta \tau, \quad \tilde{x}_{i} = \sqrt{\lambda} \eta x_{i}, \quad \tilde{\phi} = \frac{\phi}{\eta},
\end{eqnarray}
where $\tau$ denotes the physical conformal time, $x_i$ represents the comoving distance, and tilded quantities correspond to their program variable counterparts. The program potential is defined as
\begin{widetext}
\begin{eqnarray}\label{potential_pro}
    \tilde{V}(\tilde{\phi})\equiv\frac{1}{\lambda\eta^4}V(\eta\tilde{\phi})
    =\frac{1}{4} (|\tilde{\phi}|^2 - 1)^2 + 
    \left(\frac{m}{\sqrt{\lambda}\eta}\right)^2\frac{1}{N^2} (1 - |\tilde{\phi}|\cos N \theta)+\Xi_1 \tilde{\phi}_1 + \Xi_2 \tilde{\phi_2}.
\end{eqnarray}
\end{widetext}
The explicit equations of motion, expressed in terms of program variables, are provided in Appendix~\ref{EoM}.

\section{Field configurations and evolution of the area density}\label{sec:Evolution of Domain Walls}
First, we verify whether the scaling properties of domain walls hold in our simulations. In addition to the basic parameters listed in Appendix~\ref{sec:Parameters}, we set $\Xi_1=\Xi_2=0$. We compute the area parameter using the method described in Appendix~\ref{Cal of AD}. Figure~\ref{fig:scaling} presents the evolution of the area parameters over time in the absence of bias. As observed, once domain walls stabilize, they enter the scaling regime, during which the area parameters for different domain walls remain nearly constant. {The initial fluctuations or ``jumps" in the area parameters arise from the formation process of the domain wall network, during which the field configuration transitions from random initial conditions to a physically meaningful domain wall structure.} We perform a constant fit to the total area parameter, \(\mathcal{A}_{tot}\), for \(\tilde{\tau} > 16\) and obtained an approximate value of 2.223 during the scaling stage. Thus, our simulations confirm that the scaling properties of stable domain walls are preserved. {The late-stage increase in the area parameter is attributed to numerical errors~\cite{kawasaki_axion_2015}.}
\begin{figure}[ht]    \includegraphics[width=0.45\textwidth]{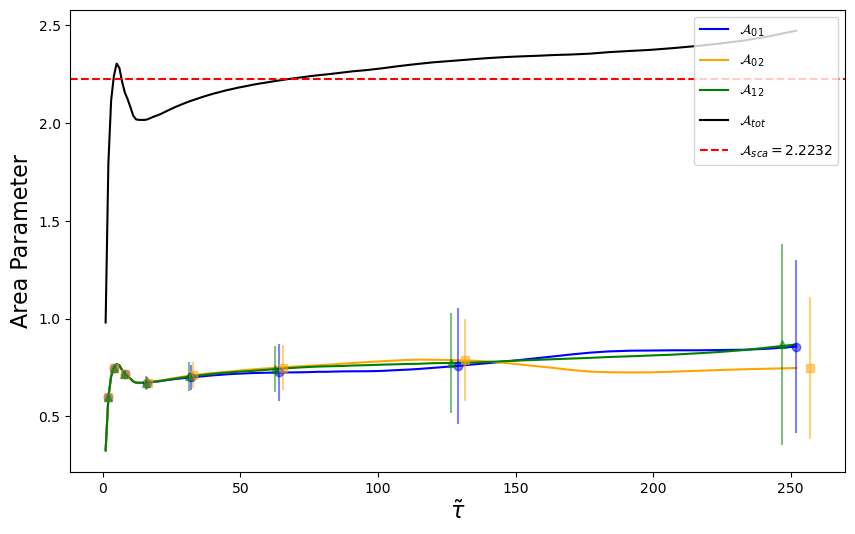}
    \caption{Evolution of the area parameters for each type of domain wall, as well as the total area parameter, with respect to the program time $\tilde{\tau}$, in the absence of bias, $\Xi_1 = \Xi_2 = 0$. {Each data point with error bars corresponds to the average over 8 independent realizations.}}
    \label{fig:scaling}
\end{figure}

\begin{figure*}[h]
\includegraphics[width=0.45\linewidth]{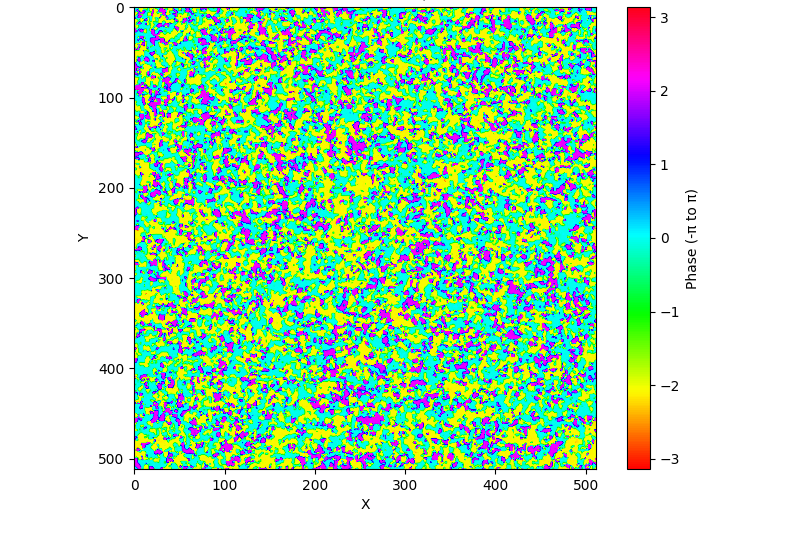}
    \includegraphics[width=0.45\linewidth]{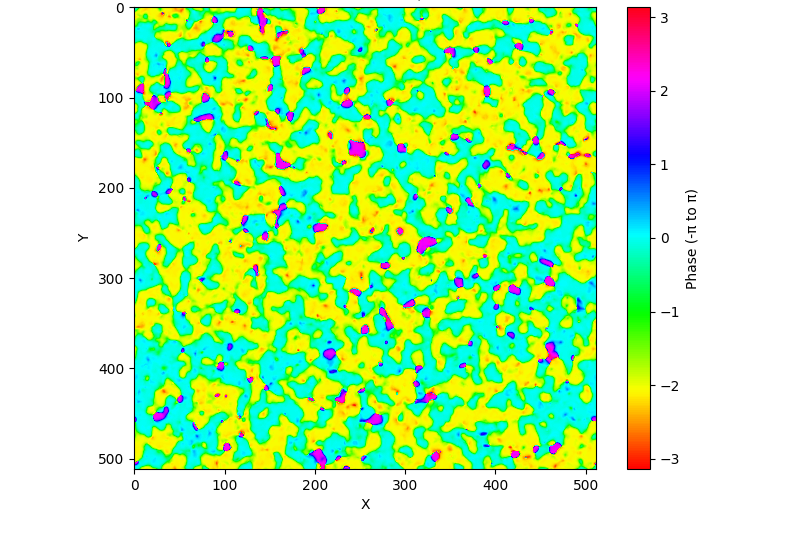}
    \includegraphics[width=0.45\linewidth]{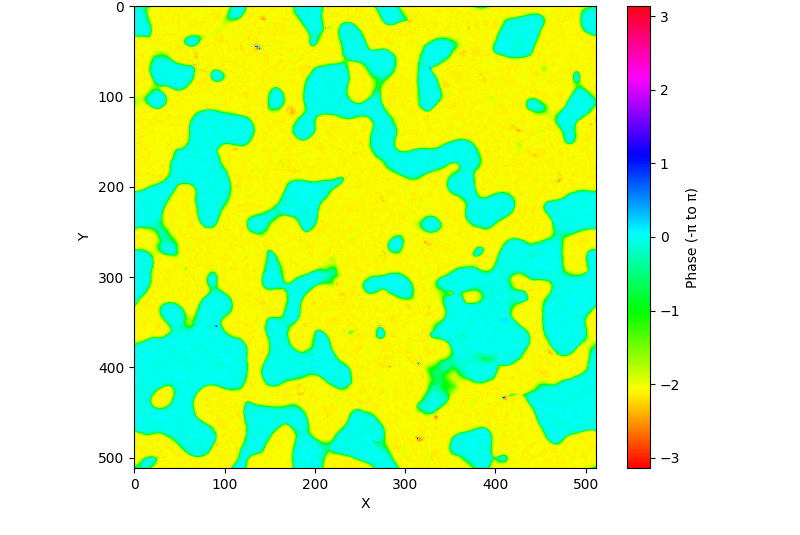}
    \includegraphics[width=0.45\linewidth]{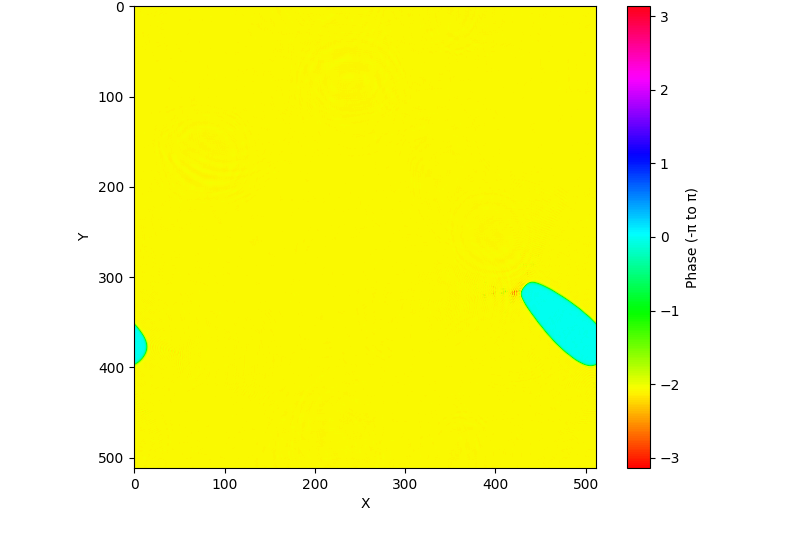}
    \caption{Phase distributions of the field at different moments, with model parameters are $\Xi_2=0.02, \Xi_1=-0.01$, are shown in slices. Arranged from top left to bottom right, these correspond to program times \(\tilde{\tau}\) of 5, 13, 37, and 256, respectively. The phase values are mapped continuously within the range of \(-\pi\) to \(\pi\), represented by a gradient of colors.}
    \label{slice}
\end{figure*}

Next, we observe the evolution of unstable domain walls by setting \(\Xi_2 = 0.02\) and \(\Xi_1 = -0.01\). The spatial distribution of the scalar field phase at different program times is shown in slices in Fig.~\ref{slice}. Initially, cosmic strings and domain walls connected to them are formed, separating different domains. Then, Vacuum 2 begins to decay, leaving only Vacuum 0 and Vacuum 1 in our simulated box. After some evolution, Vacuum 1 also begins to decay, and finally, only Vacuum 0 dominates.

\begin{figure}[ht]
    \centering
    \includegraphics[width=0.45\textwidth]{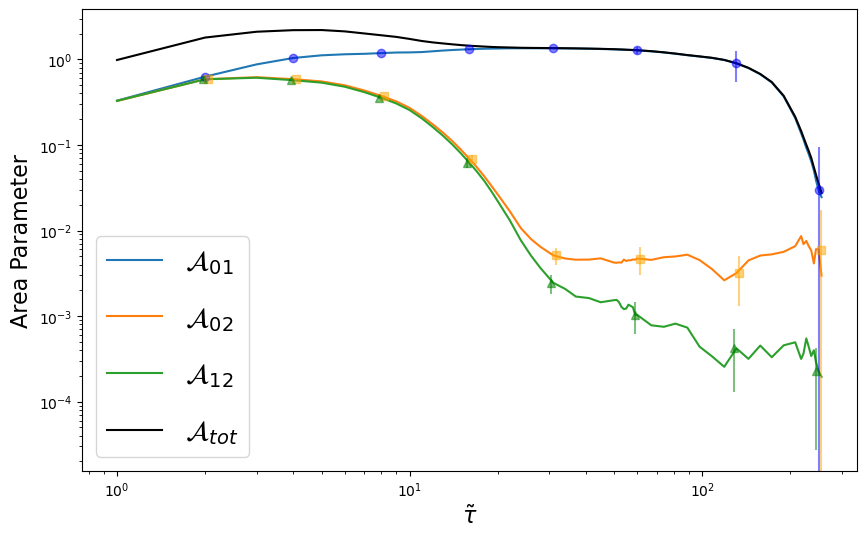}
    \caption{
Evolution of the area parameters for each type of domain wall, as well as the total area parameter, with respect to the program time $\tilde{\tau}$, when parameters are set to \(\Xi_1 = -0.01\) and \(\Xi_2 = 0.02\). {The plot shows the transition from the scaling regime to the annihilation stage, with the duration of the scaling regime varying across different domain wall types.} {Each data point with error bars corresponds to the average over 8 independent realizations.}
}
    \label{area_parameter_fig}
\end{figure}

Figure~\ref{area_parameter_fig} illustrates the evolution of the area of various domain walls over the program time \(\tilde{\tau}\) with the parameters set to \(\Xi_1 = -0.01\) and \(\Xi_2 = 0.02\), corresponding to \(\zeta \approx\ 0.066\). As we can find that the annihilation of domain walls is divided into two stages. In the first stage, the annihilation of the 12-DW and 02-DW occurs simultaneously due to the decay of the Vacuum 2. After the annihilation of these two types of domain walls, only the 01-DW remains. As the system evolves, the 01-DW also annihilates after a certain period.

Additionally, we find that as the 12-DW and 02-DW annihilate, the area of the 01-DW slightly increases. This can be interpreted as a process where the 02-DW and 12-DW 
 ``merge" into the 01-DW during the decay of Vacuum 2. Due to this process, the total area decreases more gradually during the first stage of the domain wall annihilation.

We notice that, near the moment of annihilation of domain walls, the area of domain walls oscillates due to field fluctuations. Therefore, we need to reconsider our estimation of the annihilation time, rather than simply using the moment when the area parameter reaches zero as the annihilation time. For this reason, we define the annihilation time of each type of domain walls as the time when the area parameter first drops below 0.01. This corresponds to the level where the area parameter is approximately 1\% of its value during the scaling regime.


\section{Estimation of the annihilation time}\label{sec:Estimation of the Time of Annihilation}

The fixed parameters listed in Appendix~\ref{sec:Parameters} are used in our simulations, and we set $\Xi_2 = 0.02$, which results in $\Delta \tilde{V} \approx \sqrt{3\Xi_2} = 0.0342$. {This value of the largest potential difference is adopted consistently in all simulations presented in this section. The values of $\Xi_1$ and the corresponding $\zeta$ used in our analysis are listed in Table~\ref{tab:parameter}.}

\begin{table}[ht]
    \centering
    \begin{tabular}
    {c|c|c|c|c|c|c|c|c|c|c|c}
        \hline
        \hline
        $\quad\Xi_1\quad$ & $-0.01$&$-0.008$&$-0.007$&$-0.006$&$-0.004$&$-0.002$&0&0.002&0.004&0.006&0.01\\
        \hline
        $\quad\zeta\quad$ & 0.067&0.153&0.196&0.239&0.325&0.411&0.498&0.584&0.671&0.758&0.933\\
        \hline
        \hline
    \end{tabular}
    \caption{Values of $\Xi_1$ and $\zeta$ used in simulations.}\label{tab:parameter}
\end{table}

\begin{figure}[ht]
    \centering
    \includegraphics[width=0.45\textwidth]{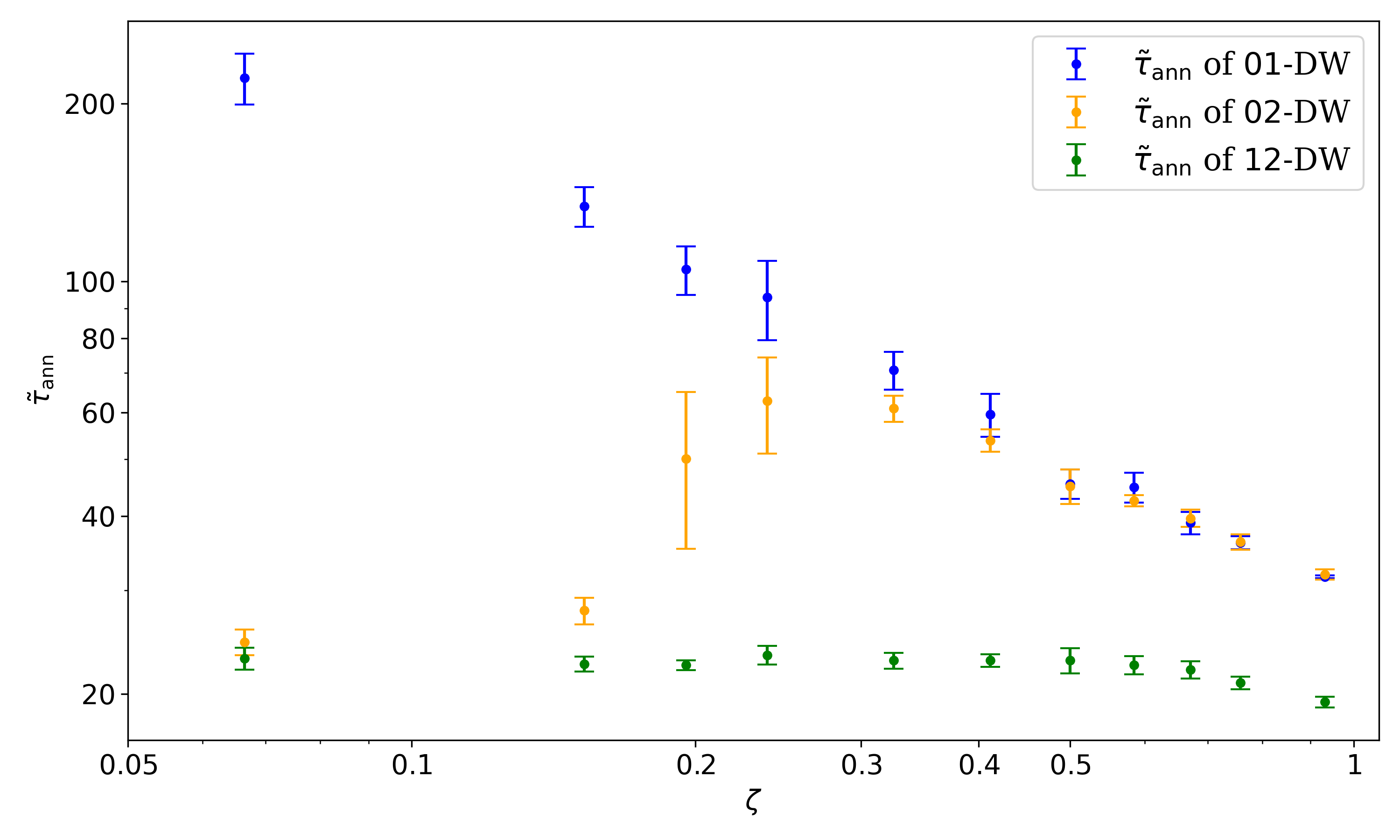}
    \caption{Dependence of the annihilation time (in program units) of various domain walls on the parameter $\zeta$. {The plot shows raw simulation results with error bars. Each data point with error bars corresponds to the average over 8 independent realizations.}}
    \label{fig:t_ann}
\end{figure}

Using the method for measuring the annihilation time proposed at the end of Sec.~\ref{sec:Evolution of Domain Walls}, {we investigated how the annihilation time of different types of domain walls depends on the parameter $\zeta$. The simulation results are shown in Fig.~\ref{fig:t_ann}.} 

Guided by the simulation results, we derive a semi-analytical expression for the annihilation time of different types of domain walls. In particular, for the 01-DW, during the late stages of evolution when only Vacuum 0 and Vacuum 1 persist, its dynamics resemble those of $Z_2$ domain walls. In the $Z_2$ case, the annihilation time is expected to scale inversely with the potential difference $\Delta V_{01}$ as
\begin{equation}\label{anylatically}
t_{\text{ann},01} \sim \frac{\sigma}{\Delta V_{01}} = \frac{8m\eta^2}{9\zeta \Delta V}.
\end{equation}
Therefore, it is natural to expect a similar inverse dependence for the 01-DW.

However, in the $N=3$ scenario considered here, the situation is more complex: the decay probability of Vacuum 2 into Vacuum 0 exceeds that into Vacuum 1, introducing an asymmetry that can influence the effective annihilation dynamics of the 01-DW. Indeed, we find that fixing the slope to the theoretical value of $-1/2$ fails to accurately describe the numerical data in the regime of relatively large $\zeta$, where the domain wall decays more rapidly and may not have fully reached the scaling regime. The influence of the multi-vacuum structure becomes non-negligible in this region, further motivating a more flexible fitting strategy.

\begin{figure}[ht]
    \centering
    \includegraphics[width=0.5\linewidth]{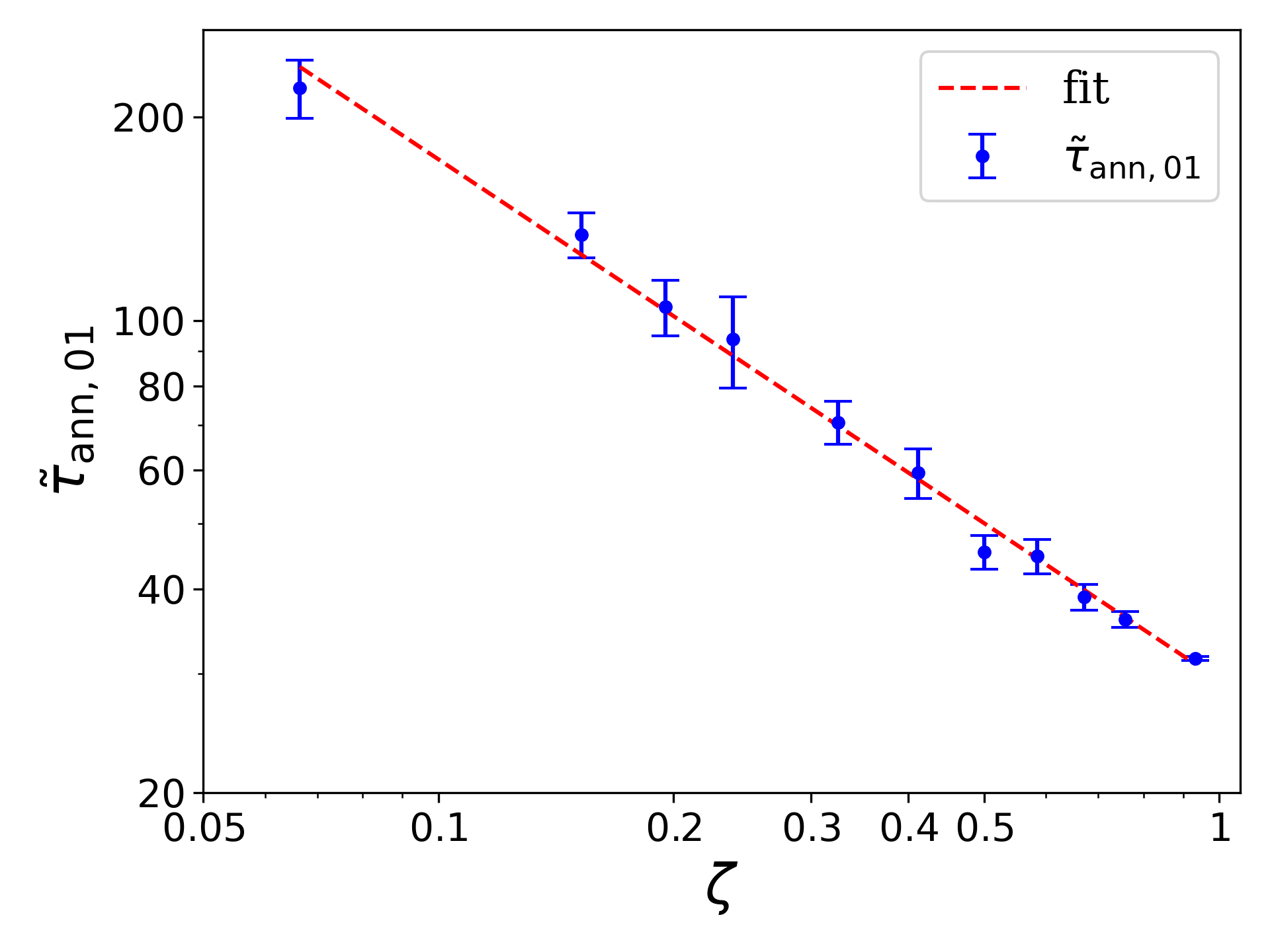}
    \caption{Annihilation time of the 01-DW, \( \tilde{\tau}_{\text{ann},01} \), as a function of the potential difference ratio \( \zeta \). The red dashed line indicates the best-fit constant.}
    \label{fig:t_ann_01}
\end{figure}

To account for these effects, we adopt a generalized power-law dependence on the potential difference, treating the exponent $a$ as a free parameter to be determined through numerical fitting. Accordingly, the dimensionless annihilation time is modeled as
\begin{eqnarray}\label{eq:tau_general}
\tilde{\tau}_{\text{ann},01} = \kappa_{01} \left( \frac{m}{\sqrt{\lambda} \eta} \right)^{1/2} \left( \zeta \Delta\tilde{V} \right)^{-a},
\end{eqnarray}
where $\kappa_{01}$ and $a$ are numerical coefficients obtained via least-squares regression on the logarithmic scale. {The best-fit analysis shown in Fig.~\ref{fig:t_ann_01} yields}

\begin{eqnarray}\label{eq:fit_result}
\kappa_{01} = 10.81 \pm 0.67, \quad a = 0.7705 \pm 0.0208,
\end{eqnarray}
and consequently,
\begin{eqnarray}\label{eq:t_ann_01}
t_{\text{ann},01} \approx 58.43 \, \frac{m\eta^2}{\left( \zeta \Delta V \right)^{1.54}}.
\end{eqnarray}
This result generalizes the conventional inverse scaling behavior observed in the $Z_2$ scenario and more accurately characterizes the annihilation dynamics in the $Z_3$ framework.

\begin{figure}[ht]
    \centering
    \includegraphics[width=0.5\linewidth]{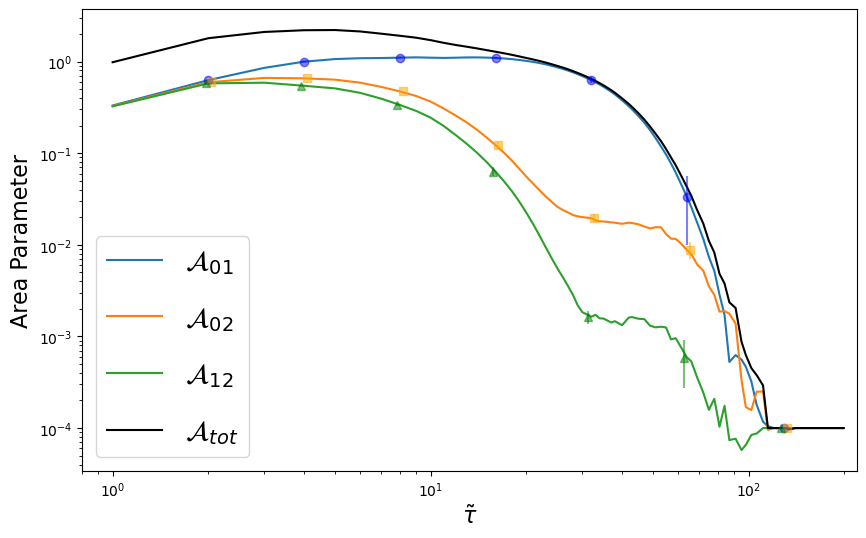}
    \caption{{Evolution of the area parameters for each type of domain wall, as well as the total area parameter, with respect to the program time $\tilde{\tau}$, when parameters are set to \(\Xi_1 = -0.004\) and \(\Xi_2 = 0.02\), corresponding to $\zeta=0.325$. In this case, the 01-domain wall annihilates shortly after the other two types of domain walls. Since there is no sufficiently long period of scaling, the analytical estimate of the domain-wall annihilation time (Eq.~\ref{anylatically}) breaks down. Each data point with error bars corresponds to the average over 8 independent realizations.}}
    \label{fig:area_parameter_zeta_0.325}
\end{figure}

\begin{figure}[ht]
    \centering
    \includegraphics[width=0.5\linewidth]{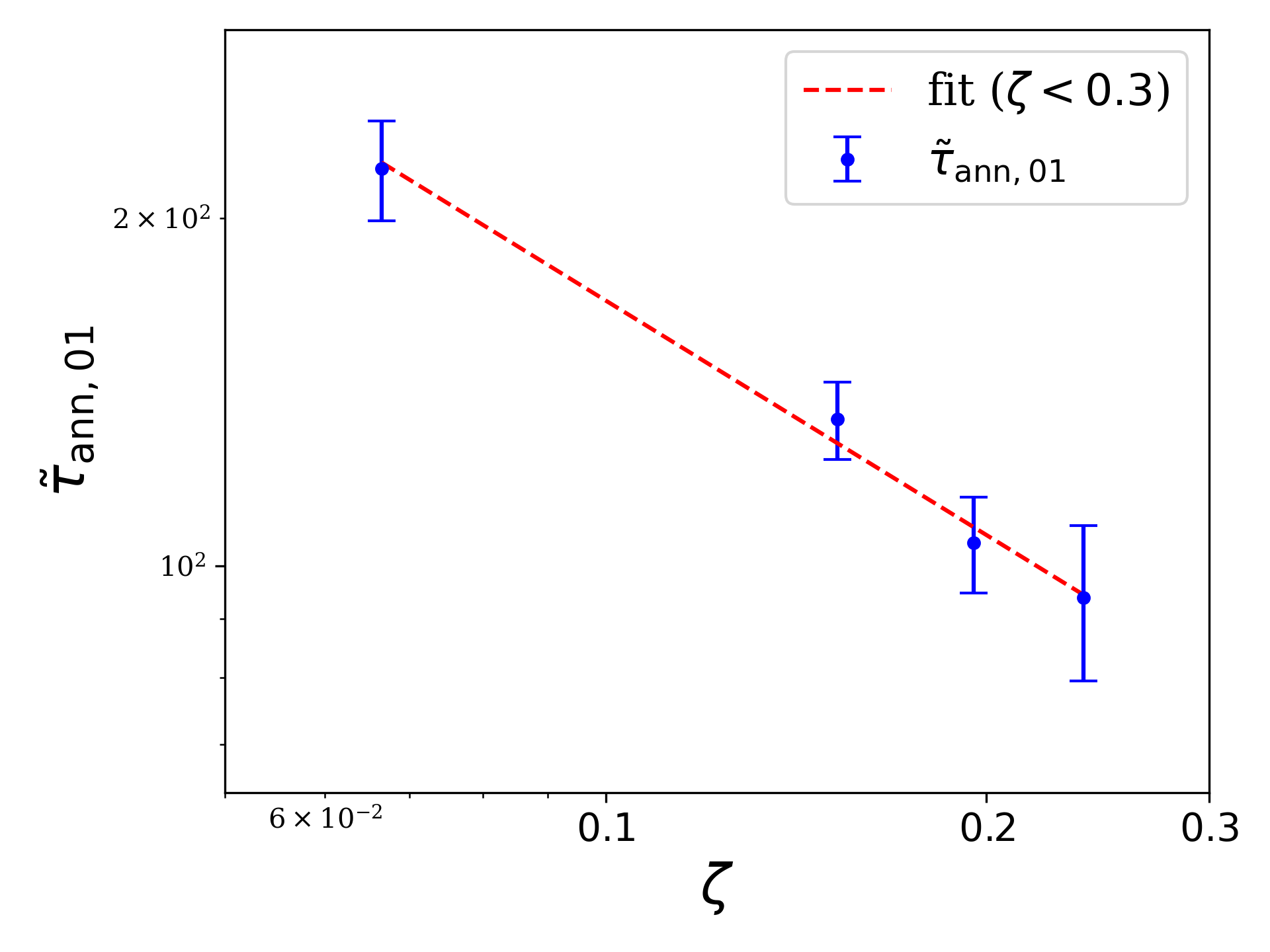}
    \caption{{Annihilation time of the 01-DW, \( \tilde{\tau}_{\text{ann},01} \), as a function of the potential difference ratio \( \zeta \). The analysis is restricted to the regime \( \zeta < 0.3 \). The red dashed line indicates the best-fit constant, and the corresponding slope is \( a = 0.6749 \pm 0.0427 \).}
}
    \label{fig:t_ann_01_zeta0.3}
\end{figure}

For relatively large $\zeta$, the domain wall decays more rapidly and may not yet have {sufficiently} reached the scaling regime, {as illustrated in Fig.~\ref{fig:area_parameter_zeta_0.325}}. The influence of the multi-vacuum structure on the decay time may also become non-negligible. This results in that the best-fit slope deviates from the theoretical value of $-1/2$. However, we found that for data points with $\zeta < 0.3$, {the numerically extracted slope agrees more closely with the theoretical value of $-1/2$, as illustrated in Fig.~\ref{fig:t_ann_01_zeta0.3}.} 

The annihilation of the 12-DW has not been explicitly discussed in previous studies. According to our simulation results, it is essentially insensitive by bias directions.

For small values of \( \zeta \), $\Delta V_{01}$ is rather small. As the vacuum bubbles of Vacuum 2 collapse, the 12-DW and the 02-DW annihilate almost simultaneously, as shown in Fig.~\ref{area_parameter_fig}. 

\begin{figure*}[h]
    \centering
    \includegraphics[width=0.45\textwidth]{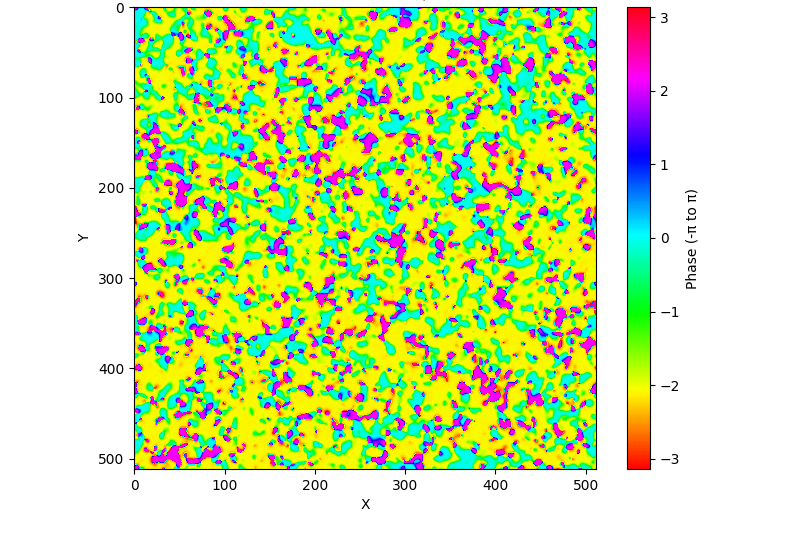}
    \includegraphics[width=0.45\textwidth]{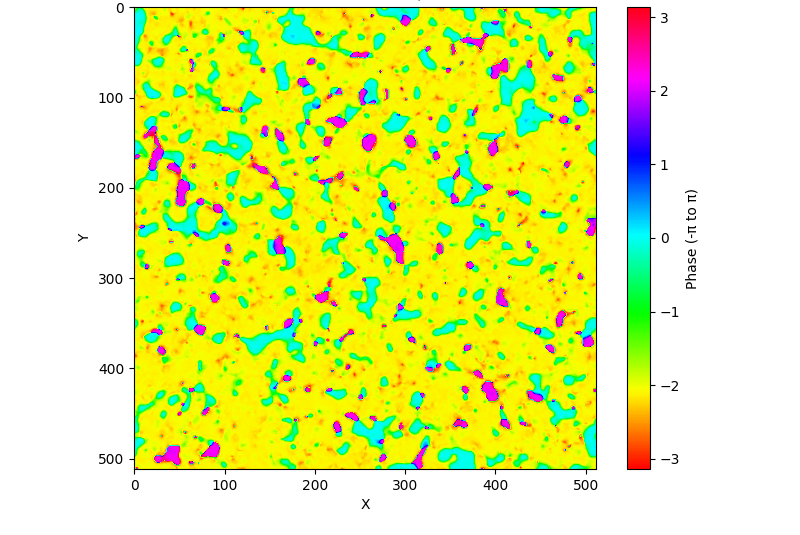}
    \includegraphics[width=0.45\textwidth]{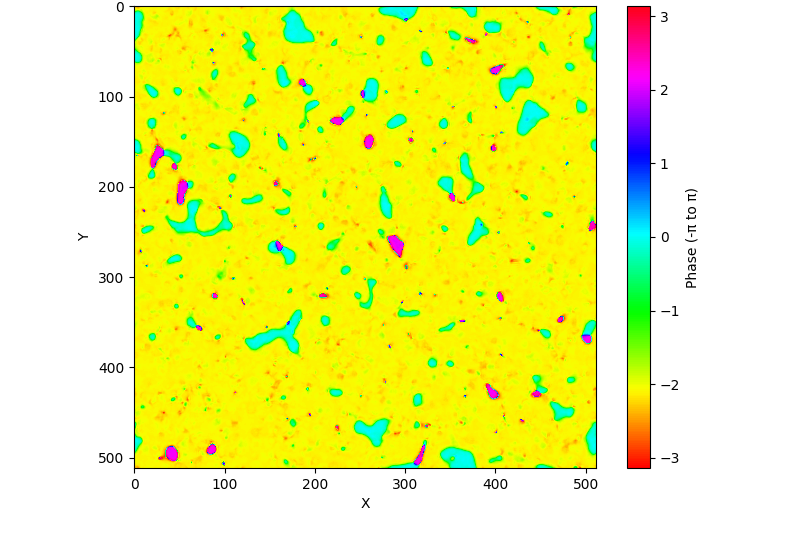}
    \includegraphics[width=0.45\textwidth]{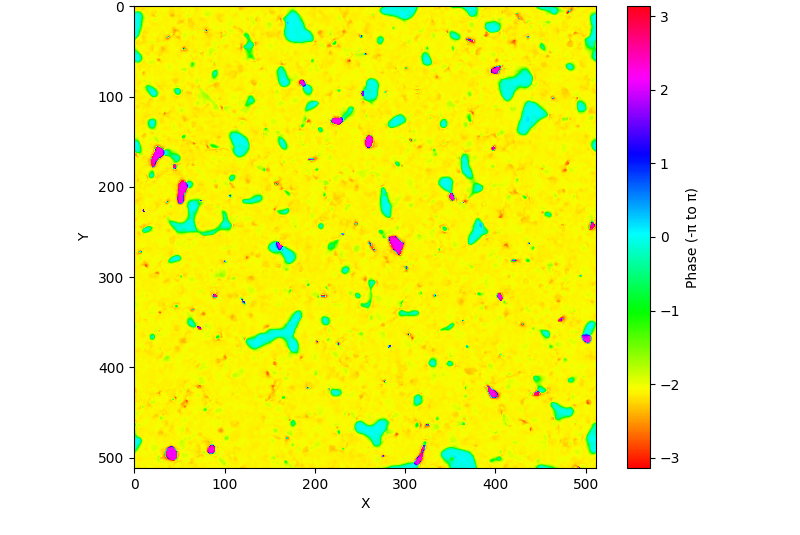}
    \caption{Phase distribution of the field at different moments for model parameters $\Xi_2=0.02, \Xi_1=0.004$, shown in slices. Arranged from top left to bottom right, these correspond to program times \(\tilde{\tau}\) of 9, 14, 19, and 20, respectively. The phase values are continuously mapped within the range of \(-\pi\) to \(\pi\), represented using a gradient of colors.}
    \label{slice_0.004}
\end{figure*}

For larger values of \( \zeta \), $\Delta V_{12}$ becomes smaller. As both Vacuum 1 and Vacuum 2 decay into Vacuum 0, the size of the vacuum bubbles of Vacuum 1 and Vacuum 2  shrink below the Hubble horizon. At this stage, the 12-DW separating the two vacua forms a finite-sized, closed string boundary. Due to the string tension, this boundary rapidly contracts, causing the 12-DW to annihilate quickly. As a result, the two vacuum bubbles on either side separate into distinct individual bubbles. In Fig.~\ref{slice_0.004}, the red bubble near \(x=400, y=80\) separating from the cyan bubble serves as an example of this process.

Therefore, for small \( \zeta \), the annihilation of the 12-DW is mainly driven by the pressure difference on either side of domain walls, whereas for large $\zeta$, it is predominantly governed by the tension of the closed string. 

{Since $t_{\text{ann},12}$ is found to be insensitive to the bias direction, we estimate the annihilation time of the 12-domain wall by
\begin{eqnarray}\label{eq:tau_12}
\tilde{\tau}_{\text{ann},12}=\kappa_{12} \left(\frac{m}{\sqrt{\lambda}\eta}\right)^{1/2}\Delta\tilde{V}^{-1/2},
\end{eqnarray}
the numerical coefficient \(\kappa_{12}\) is obtained by fitting the simulation results. From Fig.~\ref{fig:t12}, the best-fit value for the numerical coefficient $\kappa_{12}$ is found to be
\begin{eqnarray}\label{kappa_12}
    \kappa_{12}=4.11\pm0.06.
\end{eqnarray}
\begin{figure}[ht]
    \centering
    \includegraphics[width=0.5\linewidth]{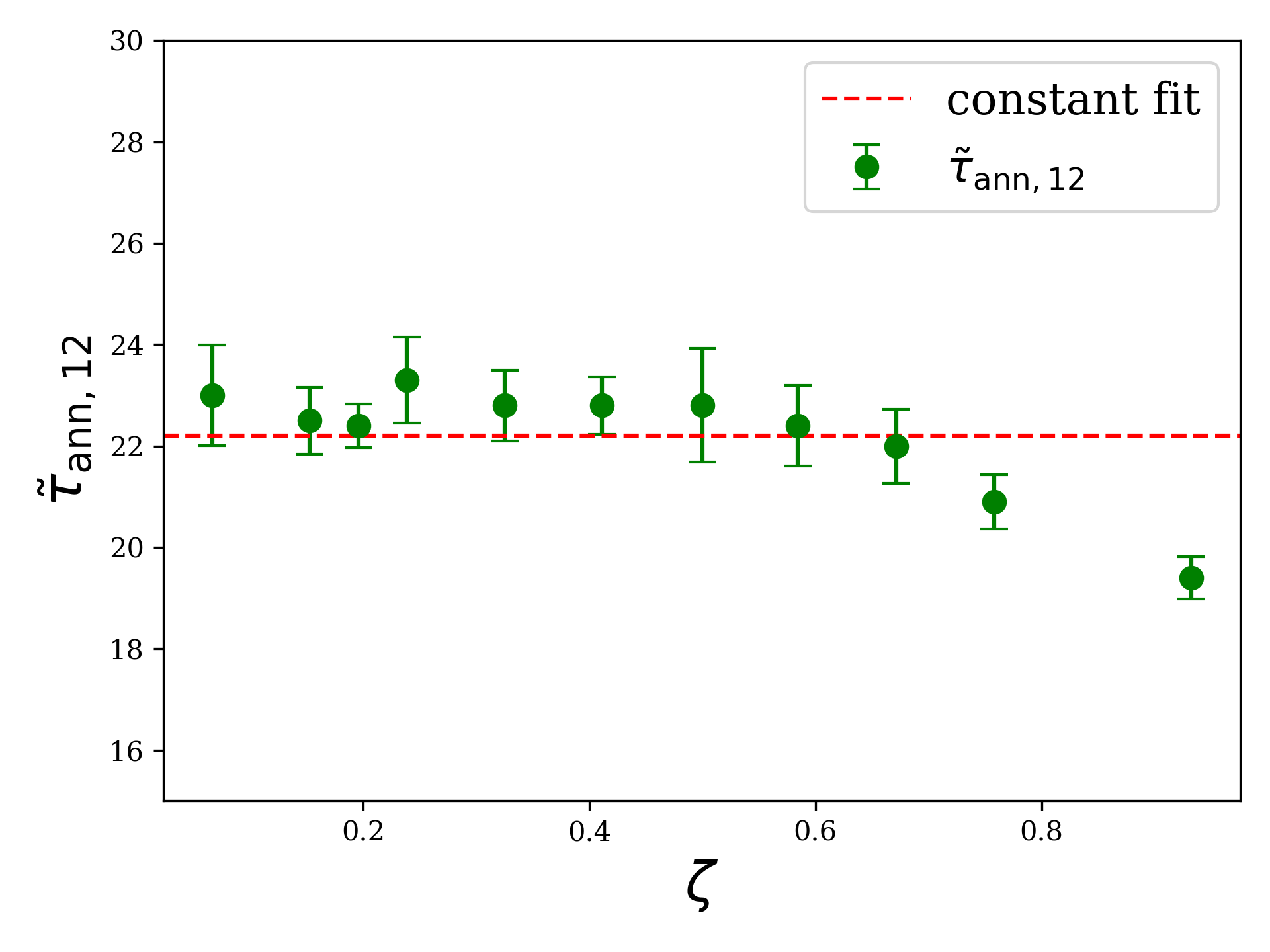}
    \caption{{Annihilation time of the 12-DW, \( \tilde{\tau}_{\text{ann},12} \), as a function of the potential difference ratio \( \zeta \). The red dashed line indicates the best-fit constant.}}
    \label{fig:t12}
\end{figure}
Therefore, the physical annihilation time of the 12-DW can be estimated as
\begin{eqnarray}\label{eq:t_12}
t_{\text{ann},12}\approx8.45\frac{m\eta^2}{\Delta V}.
\end{eqnarray}}

The dependence of the annihilation time of 02-DW $\tilde{\tau}_{\text{ann},02}$ on \(\zeta\) can be analyzed in two regimes. For \(\zeta < 0.3\), the annihilation time increases with \(\zeta\). This can be understood as follows: as the energy of Vacuum 1 increases, the potential difference between Vacuum 2 and Vacuum 1 decreases, leading to a slower decay rate from Vacuum 2 to Vacuum 1. As a result, the decay time of Vacuum 2 increases. For \(\zeta > 0.3\), the decay rate from Vacuum 1 to Vacuum~0 becomes sufficiently large, allowing Vacuum 0 to quickly dominates. After this, the remaining two vacua almost decay synchronously.

In previous research, the relationship between the annihilation time of the 02-DW and bias directions has been rarely explored. Researchers have often estimated the annihilation time of this type of domain walls using methods similar to those for $Z_2$ domain walls. However, our simulation results indicate that the annihilation time of the 02-DW is not solely determined by the potential difference between Vacuum 2 and the true vacuum; rather, it also exhibits a nontrivial dependence on bias directions. {This complexity arises from the multi-vacuum interactions involved, making the process inherently nonlinear and stochastic. As a result, no established analytical model currently exists to describe this behavior quantitatively.}

\section{Summary and discussion}\label{summary}

In this study, we have investigated the impact of bias directions on the dynamics of $Z_N$ domain walls, with a particular focus on quantitatively describing the evolution of domain walls by estimating their annihilation times. We set \(N = 3\) to examine the dynamics of domain walls in a $Z_3$ symmetric system and analyzed how strings and domain walls arise. To avoid the domain wall problem, we introduced bias terms into the potential. In addition, we used $\Delta V$ to quantify the magnitudes of the biases and $\zeta$ to characterize their directions.

We numerically investigate the nonlinear dynamics of domain walls in a radiation-dominated Universe where lattice simulations are performed in a cubic box. We employed dimensionless quantities in our calculations to ensure the results are general. Simulations explored the domain wall evolution under various bias configurations, controlled by the parameters $\Xi_1$ and $\Xi_2$, and examined their effects on annihilation times.

Based on our simulation results, we have derived analytical expressions for annihilation times of domain walls. For the 01-DW, the annihilation time $t_{\text{ann},01}$ exhibits a behavior qualitatively similar to the case of $N=2$, but follows a distinct power-law scaling with respect to the potential difference. For the 12-DW,the annihilation time is roughly unaffected by bias directions, estimated as $t_{\text{ann},12} \sim \sigma/\Delta V$. The annihilation time of the 02-DW exhibits a more intricate dependence on both the potential difference and bias directions. For $\zeta < 0.3$, $t_{\text{ann},02}$ increases with $\zeta$, while for $\zeta > 0.3$, it aligns with the behavior of the 01-DW, decreasing with $\zeta$. More detailed estimations are given by Eqs.~\eqref{eq:t_ann_01}, \eqref{eq:t_12}. 

The estimate of $t_{\text{ann},01}$ is similar with previous  results~\cite{Hiramatsu:2010yn,kawasaki_axion_2015}. The annihilation time of domain walls between the false vacua, such as $t_{\text{ann},12}$, has not been systematically investigated in previous studies. The estimate of $t_{\text{ann},02}$ corrects the previous simple assumption that it is independent of bias directions~\cite{wu2022collapsing}.

Although our study primarily focuses on the evolution and dynamics of $Z_3$ domain walls, the characterization of bias directions in this work can also be generalized to the $N>3$ cases. Further investigations can be conducted. We predict that for $N>3$, the dynamics of domain walls will become more complex, involving more independent parameters.

Stepwise annihilation of domain walls is expected to produce double-peak or multi-peak structures of the GW energy spectrum~\cite{wu2022collapsing}, leading to more distinguishable observational GW signals arising from $Z_N$ domain walls compared to single-peak structures. The peak amplitude and frequency of the GW spectrum generated by domain walls depend on their annihilation time~\cite{Kadota:2015dza,kawasaki2011study,hiramatsu2014estimation}. Therefore, our estimate of the domain wall annihilation times can improve the accuracy of predictions for the GW spectrum arising from $Z_N$ domain walls.

Additionally, we observe the remaining domain wall area temporarily increases due to the stepwise annihilation of domain walls. Consequently, the reduction in the total area parameter during each stage of the domain wall annihilation is not uniform. By fully considering this effect, we may achieve more precise estimations of the GW spectrum produced by domain walls.
Furthermore, we anticipate that future studies could directly simulate the GW spectrum to examine or reveal the dependence of its spectral shape on bias directions. Such simulations could also provide an opportunity to observe double-peak or multi-peak structures in the GW spectrum arising from $Z_N$ domain walls.

\section*{ACKNOWLEDGMENTS}
This work is supported in part by the National Key Research and Development Program of China Grant No. 2020YFC2201501, in part by the National Natural Science Foundation of China under Grant No. 12475067 and No. 12235019.

\appendix
\section{Equations of Motion}\label{EoM}

In our simulations, we define two real fields to represent a complex scalar field, \( \tilde{\phi} = \tilde{\phi}_1 + i\tilde{\phi}_2 \). Based on Eqs.~\eqref{eq:prs} and \eqref{potential_pro}, the equations of motion for the real scalar fields are given as follows:
\begin{widetext}
\begin{align}
    \tilde{\phi}_1''+3\frac{a'}{a}\tilde{\phi}_1'-\nabla^2\tilde{\phi}_1
    &= -\Xi_1-\tilde\phi_{{1}}\left(-1+\tilde\phi_{\mathbf{1}}^{2}+\tilde\phi_{2}^{2}\right) 
    +\frac{m^{2}}{{N^{2}\sqrt{\tilde\phi_{1}^{2}+\tilde\phi_{2}^{2}}}}\left[\cos\left(N\theta\right)\tilde\phi_{1}+N\sin\left(N\theta\right)\tilde\phi_{2}\right],\\
    \tilde{\phi}_2''+3\frac{a'}{a}\tilde{\phi}_2'-\nabla^2\tilde{\phi}_2
    &= -\Xi_2-\tilde\phi_{{2}}\left(-1+\tilde\phi_{\mathbf{1}}^{2}+\tilde\phi_{2}^{2}\right) 
    -\frac{m^{2}}{{N^{2}\sqrt{\tilde\phi_{1}^{2}+\tilde\phi_{2}^{2}}}}\left[\cos\left(N\theta\right)\tilde\phi_{2}+N\sin\left(N\theta\right)\tilde\phi_{1}\right].
\end{align}
\end{widetext}
Here, a prime denotes differentiation with respect to the program time, and \( \nabla^2 \) represents the Laplacian operator with respect to the program coordinates.

\section{Initial Conditions}\label{IC}
The scale factor can be expressed in terms of the dimensionless conformal time as
\begin{eqnarray}\label{atau}
    a(\tilde{\tau}) = \frac{\sqrt{\lambda} \eta}{H_i} \frac{\tilde{\tau}}{\tilde{\tau}_i^2},
\end{eqnarray}
where $H_i$ is the initial Hubble parameter. By convention, we set the initial scale factor $a(\tilde{\tau_i})$ to unity,
\begin{eqnarray}\label{ataui}
    a(\tilde{\tau}_i) = 1.
\end{eqnarray}
Although the initial time in simulations is arbitrary, we choose
\begin{eqnarray}
    H_i = \sqrt{\lambda} \eta,
\end{eqnarray}
which simplifies the initial dimensionless conformal time to
\begin{eqnarray}
    \tilde{\tau}_i = 1.
\end{eqnarray}
Thus, we derive the simple relation,
\begin{eqnarray}
    \tilde{\mathcal{H}}_i^{-1}(\tilde{\tau})=a({\tilde{\tau}})=\tilde{\tau},
\end{eqnarray}
where $\tilde{\mathcal{H}}$ denotes the dimensionless comoving Hubble parameter, and $\tilde{\mathcal{H}}_i^{-1}$ represents the dimensionless comoving Hubble horizon.

Since the late stage evolution of domain walls is largely independent of the initial conditions~\cite{dankovsky2024revisiting}, we adopt a Gaussian distribution for simplicity. The initial mean values of the scalar field \( \tilde{\phi}_i \) and its time derivative \( \tilde{\phi}'_i \) are set to zero, while their perturbations are characterized in momentum space by the correlation functions,

\begin{align}
\langle \tilde{\phi}_i(\mathbf{k}) \tilde{\phi}_i(\mathbf{k}') \rangle &= \frac{1}{2k} (2\pi)^3 \delta^{(3)}(\mathbf{k} + \mathbf{k}'), \\
\langle \tilde{\phi}'_i(\mathbf{k}) \tilde{\phi}'_i(\mathbf{k}') \rangle &= \frac{k}{2} (2\pi)^3 \delta^{(3)}(\mathbf{k} + \mathbf{k}'), \quad (i = 1, 2).
\end{align}
Here, \( \tilde{\phi}' \) represents the time derivative of the scalar field in simulations, \( \mathbf{k}' \) refers to modes distinct from \( \mathbf{k} \), and \( k \) represents the magnitude of \( \mathbf{k} \). Since the effective mass of the scalar field is initially negative, we approximate the field as massless by replacing \( \sqrt{k^2 + m^2} \) with \( k \). 

To suppress unphysical noise from high-frequency modes, we introduce a momentum cutoff \( k_{\text{cut}} \). Fluctuations are set to zero for all modes with \( k > k_{\text{cut}} \). In our setup, we set the cutoff value to \( \tilde{k}_{\text{cut}} = 3 \).

\section{Parameters for simulations}\label{sec:Parameters}

Since simulations are performed with dimensionless quantities, the simulation results are independent of the specific values of the dimensionless coefficients. This allows us to set these parameters arbitrarily. We set \( \lambda = 0.1 \) and \( \eta = 2 \times 10^{17} \, \text{GeV} \). Additionally, We choose \( m = \sqrt{\lambda} \eta \) for clarity. 

Simulating topological defects presents two primary challenges: \romannumeral 1) To ensure resolvability, the defect thickness must be larger than \(L/\mathrm{N}\), the ultraviolet limit.\romannumeral 2) The Hubble horizon must remain smaller than the box size, defining the infrared limit.

The physical thicknesses are given by \(\delta_s \approx (\sqrt{\lambda} \eta)^{-1}\) for strings and \(\delta_w \approx m^{-1}\) for domain walls~\cite{Vilenkin:2000jqa}. In the PRS method, those correspond to the comoving thicknesses, \(\delta_{s,c} \approx (\sqrt{\lambda} \eta)^{-1}\) and \(\delta_{w,c} \approx m^{-1}\). The ratio of the two thicknesses is \(\delta_s / \delta_w \approx m / (\sqrt{\lambda} \eta)\). For the parameters we selected, \(\delta_w \approx \delta_s\). We choose \(\delta_{s(w),c} \approx L/\mathrm{N}\), or equivalently, in terms of mensionless quantities
\begin{eqnarray}\label{UV}
    \tilde{\delta}_{s(w),c} \approx \tilde{L}/\mathrm{N} .
\end{eqnarray}
This resolution marginally meets the requirement.

Another constraint comes from the requirement that the Hubble horizon at the end of simulations must remain within the box size, \(H_f^{-1} \leq a_f L\), or equivalently, \(\mathcal{H}_f^{-1} \leq L\), to ensure that the evolution is not influenced by boundary conditions. In terms of program variables, this condition becomes:
\begin{eqnarray}
    \tilde{\mathcal{H}}_f^{-1} = \tilde{\tau}_f \leq \tilde{L}.
\end{eqnarray}

In our simulations, \(\mathrm{N}\) is set to 512. Based on Eq.~\eqref{UV}, we choose \(\tilde{L} = 512\). Our simulations ends when the Hubble horizon reaches half the size of the comoving box, \(\tilde{\tau}_f = \tilde{\mathcal{H}}_f^{-1} = 256\).

\section{Calculation of the Area Parameter}\label{Cal of AD}
We adopt the method described in~\cite{Scherrer:1997sq, Garagounis:2002kt} to measure the comoving area of domain walls. Unlike previous studies, which computed the comoving area and area parameters by considering all domain wall types collectively, we aim to examine the evolution of each specific type of domain walls separately. Below, we outline the method used to compute the area of domain walls.

First, we divide the field values into three regions based on the phase of the saddle points in the potential function in Eq.~\eqref{potential_1} (for the case of \( \Xi = 0 \), these phases are \( \pi/3, \pi, 5\pi/3 \)). Each region contains a vacuum state and is labeled by its corresponding vacuum number. After extracting the phase for each field point at every output snapshot, we assign labels based on the corresponding field value range. See Fig.~\ref{fig:range} as an example. Hereafter, we refer to the field values within each region \(i\) as Vacuum \(i\).
\begin{figure}[ht]
    \centering
        \ctikzfig{range}
    \caption{An example of phase separation, with the potential barrier as the boundary, in the case where \(\Xi_1 = 0\) and \(\Xi_2 = 0\).}
    \label{fig:range}
\end{figure}

Next, for each field point, we compare the labels of neighboring points. If the labels of two points differ, say label \(i\) and label \(j\), we increment \( N_{ij} \) by one, where \( N_{ij} \) represents the count of the domain wall area elements corresponding to the \( ij \)-DW. This comparison is conducted along the three coordinate axes. Specifically, each grid point \((x, y, z)\) is compared with its neighbors \((x+1, y, z)\), \((x, y+1, z)\), and \((x, y, z+1)\). This procedure is applied across all grid points at each time step to determine the area count for each type of domain wall. An example of identifying a domain wall in two dimensions is presented in Fig.~\ref{fig:Identification}.
\begin{figure}[ht]
    \centering
        \ctikzfig{domain}
    \caption{An example of identifying a domain wall in two dimensions is shown. The point \((x, y)\) is labeled as Vacuum \(i\), while its two adjacent points in directions of increasing coordinates are labeled as Vacuum \(j\). From this, we infer that a domain wall spans between \((x, y)\) and its two neighboring points. Thus, we represent the area of domain walls as the line segment connecting \((x, y)\) and its two adjacent points. However, this approach overestimates the domain wall's area. To obtain the correct comoving area of domain walls, a factor of \( 2/3 \) must be applied.}
    \label{fig:Identification}
\end{figure}

As noted in~\cite{Scherrer:1997sq}, direct calculation of the domain wall area using grid points overestimates the actual comoving area. Therefore, considering the random orientations of domain walls, we introduce a correction factor of \( 2/3 \). Hence, the comoving area of each type of domain walls is estimated as:
\begin{eqnarray}
    A_{ij}(t) = \frac{2}{3} N_{ij}(t) \times \Delta x^2,
\end{eqnarray}
where \( \Delta x \equiv L/\mathrm{N} \) is the comoving lattice spacing. The area parameter is then estimated as:
\begin{eqnarray}
    \mathcal{A}_{ij} = \frac{A_{ij}(t) \tau}{V} = \frac{2}{3} \frac{N_{ij}(t) \tau}{\Delta x}.
\end{eqnarray}

In~\cite{Hiramatsu:2010yz,Li:2023gil}, the method for calculating the area of domain walls is expressed as:
\begin{eqnarray}
    A/V=C\sum_\mathrm{links}\delta\frac{|\nabla\theta|}{|\theta_{,x}|+|\theta_{,y}|+|\theta_{,z}|},
\end{eqnarray}
where $\delta$ takes the value of 1 at grid points adjacent to domain walls and 0 elsewhere. The constant $C$ is determined such that $A/V=1$. Our estimation of the domain wall area aligns with this approach in that it averages over different domain wall orientations. However, our method distinguishes between different types of domain walls, providing an additional level of differentiation.

%


\section{Lattice Artifacts}
{Our results are primarily based on simulations with a grid size of $\mathrm{N} = 512$. To examine potential lattice artifacts due to finite resolution, we also perform simulations with a larger grid size of $\mathrm{N} = 1024$, while keeping $\tilde{L}=512$ fixed, and compare the evolution of area parameters for each type of domain wall.}

\begin{figure}[ht]
    \centering
    \includegraphics[width=0.5\linewidth]{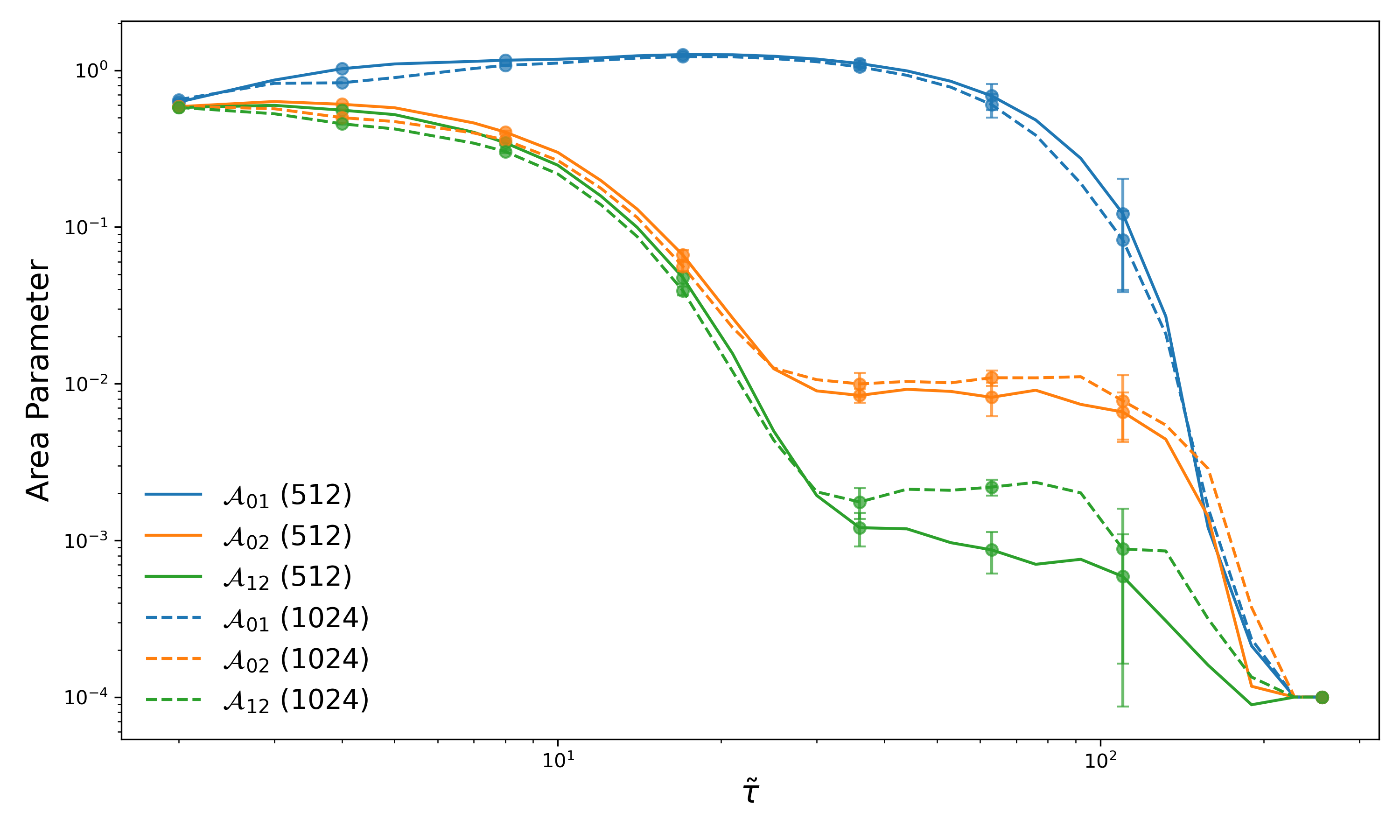}
    \caption{Evolution of the area parameters for each type of domain wall as a function of program time $\tilde{\tau}$, under different grid sizes $\mathrm{N}$. To capture the full evolution until the field oscillations subside, the parameters are set to $\Xi_1 = -0.008$ and $\Xi_2 = 0.02$. Each data point with error bars represents the average over 8 independent realizations. Solid lines correspond to $\mathrm{N} = 512$, and dashed lines to $\mathrm{N} = 1024$.
}
    \label{fig:1024}
\end{figure}

{Fig.~\ref{fig:1024} shows the evolution of the area parameters for each domain wall type with $\Xi_1 = -0.008$ and $\Xi_2 = 0.02$, under two different resolutions: $\mathrm{N} = 512$ and $\mathrm{N} = 1024$. Each data point represents an average over 8 independent realizations, with error bars indicating statistical fluctuations.
}

We find that the evolution of the area parameters before the annihilation of domain walls is nearly identical between $\mathrm{N} = 512$ and $\mathrm{N} = 1024$, with differences smaller than the statistical uncertainty. After domain wall annihilation, during the stage dominated by field oscillations, simulations with $\mathrm{N} = 1024$ exhibit slightly higher area parameters. This indicates that higher spatial resolution better captures small-scale oscillatory behavior of the field. However, under the resolution adopted in our main simulations ($\mathrm{N} = 512$), grid effects do not significantly affect the large-scale dynamics of domain walls, which are the primary focus of this study.

\bibliography{ref}

@article{dine1981simple,
    author = "Dine, Michael and Fischler, Willy and Srednicki, Mark",
    title = "{A Simple Solution to the Strong CP Problem with a Harmless Axion}",
    reportNumber = "Print-81-0320 (IAS,PRINCETON)",
    doi = "10.1016/0370-2693(81)90590-6",
    journal = "Phys. Lett. B",
    volume = "104",
    pages = "199--202",
    year = "1981"
}

@article{kibble1976topology,
    author = "Kibble, T. W. B.",
    title = "{Topology of Cosmic Domains and Strings}",
    reportNumber = "ICTP/75/5",
    doi = "10.1088/0305-4470/9/8/029",
    journal = "J. Phys. A",
    volume = "9",
    pages = "1387--1398",
    year = "1976"
}

@article{Peccei:1977ur,
    author = "Peccei, R. D. and Quinn, Helen R.",
    title = "{Constraints Imposed by CP Conservation in the Presence of Instantons}",
    reportNumber = "ITP-572-STANFORD",
    doi = "10.1103/PhysRevD.16.1791",
    journal = "Phys. Rev. D",
    volume = "16",
    pages = "1791--1797",
    year = "1977"
}

@article{Peccei:1977hh,
    author = "Peccei, R. D. and Quinn, Helen R.",
    title = "{CP Conservation in the Presence of Instantons}",
    reportNumber = "ITP-568-STANFORD",
    doi = "10.1103/PhysRevLett.38.1440",
    journal = "Phys. Rev. Lett.",
    volume = "38",
    pages = "1440--1443",
    year = "1977"
}

@article{Sikivie:1982qv,
    author = "Sikivie, P.",
    title = "{Of Axions, Domain Walls and the Early Universe}",
    reportNumber = "UFTP-82-3",
    doi = "10.1103/PhysRevLett.48.1156",
    journal = "Phys. Rev. Lett.",
    volume = "48",
    pages = "1156--1159",
    year = "1982"
}

@article{Wu:2022tpe,
    author = "Wu, Yongcheng and Xie, Ke-Pan and Zhou, Ye-Ling",
    title = "{Classification of Abelian domain walls}",
    eprint = "2205.11529",
    archivePrefix = "arXiv",
    primaryClass = "hep-ph",
    doi = "10.1103/PhysRevD.106.075019",
    journal = "Phys. Rev. D",
    volume = "106",
    number = "7",
    pages = "075019",
    year = "2022"
}

@article{wu2022collapsing,
    author = "Wu, Yongcheng and Xie, Ke-Pan and Zhou, Ye-Ling",
    title = "{Collapsing domain walls beyond Z2}",
    eprint = "2204.04374",
    archivePrefix = "arXiv",
    primaryClass = "hep-ph",
    doi = "10.1103/PhysRevD.105.095013",
    journal = "Phys. Rev. D",
    volume = "105",
    number = "9",
    pages = "095013",
    year = "2022"
}

@article{Zhitnitsky:1980tq,
    author = "Zhitnitsky, A. R.",
    title = "{On Possible Suppression of the Axion Hadron Interactions. (In Russian)}",
    journal = "Sov. J. Nucl. Phys.",
    volume = "31",
    pages = "260",
    year = "1980"
}

@article{Wilczek:1977pj,
    author = "Wilczek, Frank",
    title = "{Problem of Strong  $P$  and  $T$  Invariance in the Presence of Instantons}",
    reportNumber = "Print-77-0939 (COLUMBIA)",
    doi = "10.1103/PhysRevLett.40.279",
    journal = "Phys. Rev. Lett.",
    volume = "40",
    pages = "279--282",
    year = "1978"
}

@article{Hiramatsu:2010yn,
    author = "Hiramatsu, Takashi and Kawasaki, Masahiro and Saikawa, Ken'ichi",
    title = "{Evolution of String-Wall Networks and Axionic Domain Wall Problem}",
    eprint = "1012.4558",
    archivePrefix = "arXiv",
    primaryClass = "astro-ph.CO",
    reportNumber = "ICRR-REPORT-577-2010-10, IPMU10-0221, YITP-10-110",
    doi = "10.1088/1475-7516/2011/08/030",
    journal = "JCAP",
    volume = "08",
    pages = "030",
    year = "2011"
}

@article{Kawasaki:2014sqa,
    author = "Kawasaki, Masahiro and Saikawa, Ken'ichi and Sekiguchi, Toyokazu",
    title = "{Axion dark matter from topological defects}",
    eprint = "1412.0789",
    archivePrefix = "arXiv",
    primaryClass = "hep-ph",
    reportNumber = "ICRR-REPORT-696-2014-22, IPMU14-0348",
    doi = "10.1103/PhysRevD.91.065014",
    journal = "Phys. Rev. D",
    volume = "91",
    number = "6",
    pages = "065014",
    year = "2015"
}

@article{Weinberg:1977ma,
    author = "Weinberg, Steven",
    title = "{A New Light Boson?}",
    reportNumber = "HUTP-77/A074",
    doi = "10.1103/PhysRevLett.40.223",
    journal = "Phys. Rev. Lett.",
    volume = "40",
    pages = "223--226",
    year = "1978"
}

@article{Abbott:1982af,
    author = "Abbott, L. F. and Sikivie, P.",
    editor = "Srednicki, M. A.",
    title = "{A Cosmological Bound on the Invisible Axion}",
    reportNumber = "PRINT-82-0695 (BRANDEIS)",
    doi = "10.1016/0370-2693(83)90638-X",
    journal = "Phys. Lett. B",
    volume = "120",
    pages = "133--136",
    year = "1983"
}

@article{Dine:1982ah,
    author = "Dine, Michael and Fischler, Willy",
    editor = "Srednicki, M. A.",
    title = "{The Not So Harmless Axion}",
    reportNumber = "UPR-0201T",
    doi = "10.1016/0370-2693(83)90639-1",
    journal = "Phys. Lett. B",
    volume = "120",
    pages = "137--141",
    year = "1983"
}

@article{Preskill:1982cy,
    author = "Preskill, John and Wise, Mark B. and Wilczek, Frank",
    editor = "Srednicki, M. A.",
    title = "{Cosmology of the Invisible Axion}",
    reportNumber = "HUTP-82-A048, NSF-ITP-82-103",
    doi = "10.1016/0370-2693(83)90637-8",
    journal = "Phys. Lett. B",
    volume = "120",
    pages = "127--132",
    year = "1983"
}

@article{Press:1989yh,
    author = "Press, William H. and Ryden, Barbara S. and Spergel, David N.",
    title = "{Dynamical Evolution of Domain Walls in an Expanding Universe}",
    reportNumber = "NSF-ITP-89-51, CFA-1870",
    doi = "10.1086/168151",
    journal = "Astrophys. J.",
    volume = "347",
    pages = "590--604",
    year = "1989"
}

@article{Sousa:2010zza,
    author = "Sousa, L. and Avelino, P. P.",
    title = "{Evolution of domain wall networks: The Press-Ryden-Spergel algorithm}",
    eprint = "1101.3350",
    archivePrefix = "arXiv",
    primaryClass = "hep-th",
    doi = "10.1103/PhysRevD.81.087305",
    journal = "Phys. Rev. D",
    volume = "81",
    pages = "087305",
    year = "2010"
}

@article{Correia:2014kqa,
    author = "Correia, J. R. C. C. C. and Leite, I. S. C. R. and Martins, C. J. A. P.",
    title = "{Effects of Biases in Domain Wall Network Evolution}",
    eprint = "1407.3905",
    archivePrefix = "arXiv",
    primaryClass = "hep-ph",
    doi = "10.1103/PhysRevD.90.023521",
    journal = "Phys. Rev. D",
    volume = "90",
    number = "2",
    pages = "023521",
    year = "2014"
}

@article{Correia:2018tty,
    author = "Correia, J. R. C. C. C. and Leite, I. S. C. R. and Martins, C. J. A. P.",
    title = "{Effects of biases in domain wall network evolution. II. Quantitative analysis}",
    eprint = "1804.10761",
    archivePrefix = "arXiv",
    primaryClass = "astro-ph.CO",
    doi = "10.1103/PhysRevD.97.083521",
    journal = "Phys. Rev. D",
    volume = "97",
    number = "8",
    pages = "083521",
    year = "2018"
}

@article{Coulson:1995nv,
    author = "Coulson, D. and Lalak, Z. and Ovrut, Burt A.",
    title = "{Biased domain walls}",
    reportNumber = "UPR-0668-T",
    doi = "10.1103/PhysRevD.53.4237",
    journal = "Phys. Rev. D",
    volume = "53",
    pages = "4237--4246",
    year = "1996"
}

@article{Leite:2011sc,
    author = "Leite, A. M. M. and Martins, C. J. A. P.",
    title = "{Scaling Properties of Domain Wall Networks}",
    eprint = "1110.3486",
    archivePrefix = "arXiv",
    primaryClass = "hep-ph",
    doi = "10.1103/PhysRevD.84.103523",
    journal = "Phys. Rev. D",
    volume = "84",
    pages = "103523",
    year = "2011"
}

@article{Krajewski:2021jje,
    author = "Krajewski, Tomasz and Kwapisz, Jan Henryk and Lalak, Zygmunt and Lewicki, Marek",
    title = "{Stability of domain walls in models with asymmetric potentials}",
    eprint = "2103.03225",
    archivePrefix = "arXiv",
    primaryClass = "astro-ph.CO",
    doi = "10.1103/PhysRevD.104.123522",
    journal = "Phys. Rev. D",
    volume = "104",
    number = "12",
    pages = "123522",
    year = "2021"
}

@book{Vilenkin:2000jqa,
    author = "Vilenkin, A. and Shellard, E. P. S.",
    title = "{Cosmic Strings and Other Topological Defects}",
    isbn = "978-0-521-65476-0",
    publisher = "Cambridge University Press",
    month = "7",
    year = "2000"
}

@article{saikawa2017review,
    author = "Saikawa, Ken'ichi",
    title = "{A review of gravitational waves from cosmic domain walls}",
    eprint = "1703.02576",
    archivePrefix = "arXiv",
    primaryClass = "hep-ph",
    reportNumber = "DESY-17-036",
    doi = "10.3390/universe3020040",
    journal = "Universe",
    volume = "3",
    number = "2",
    pages = "40",
    year = "2017"
}

@article{Kadota:2015dza,
    author = "Kadota, Kenji and Kawasaki, Masahiro and Saikawa, Ken'ichi",
    title = "{Gravitational waves from domain walls in the next-to-minimal supersymmetric standard model}",
    eprint = "1503.06998",
    archivePrefix = "arXiv",
    primaryClass = "hep-ph",
    reportNumber = "CTPU-15-03, IPMU15-0031",
    doi = "10.1088/1475-7516/2015/10/041",
    journal = "JCAP",
    volume = "10",
    pages = "041",
    year = "2015"
}

@article{Marsh:2015xka,
    author = "Marsh, David J. E.",
    title = "{Axion Cosmology}",
    eprint = "1510.07633",
    archivePrefix = "arXiv",
    primaryClass = "astro-ph.CO",
    reportNumber = "KCL-PH-TH-2015-50",
    doi = "10.1016/j.physrep.2016.06.005",
    journal = "Phys. Rept.",
    volume = "643",
    pages = "1--79",
    year = "2016"
}

@article{linde1994hybrid,
    author = "Linde, Andrei D.",
    title = "{Hybrid inflation}",
    eprint = "astro-ph/9307002",
    archivePrefix = "arXiv",
    reportNumber = "SU-ITP-93-17",
    doi = "10.1103/PhysRevD.49.748",
    journal = "Phys. Rev. D",
    volume = "49",
    pages = "748--754",
    year = "1994"
}

@article{zel1974cosmological,
  title={Cosmological consequences of spontaneous violation of discrete symmetry},
  author={Zel'dovich, Yakov Boris and Kobzarev, I Yu and Okun, Lev Borisovich},
  journal={Zh. Eksp. Teor. Fiz.},
  volume={40},
  pages={3--11},
  year={1974}
}

@article{Srivastava:1999br,
    author = "Srivastava, A. M.",
    editor = "Bharadwaj, S. and Kar, S.",
    title = "{Topological defects in cosmology}",
    doi = "10.1007/s12043-999-0064-1",
    journal = "Pramana",
    volume = "53",
    pages = "1069--1076",
    year = "1999"
}

@article{Garagounis:2002kt,
    author = "Garagounis, Theodore and Hindmarsh, Mark",
    title = "{Scaling in numerical simulations of domain walls}",
    eprint = "hep-ph/0212359",
    archivePrefix = "arXiv",
    reportNumber = "SUSX-TH-02-029",
    doi = "10.1103/PhysRevD.68.103506",
    journal = "Phys. Rev. D",
    volume = "68",
    pages = "103506",
    year = "2003"
}

@article{Avelino:2005pe,
    author = "Avelino, P. P. and Oliveira, J. C. R. E. and Martins, C. J. A. P.",
    title = "{Understanding domain wall network evolution}",
    eprint = "hep-th/0503226",
    archivePrefix = "arXiv",
    doi = "10.1016/j.physletb.2005.02.003",
    journal = "Phys. Lett. B",
    volume = "610",
    pages = "1--8",
    year = "2005"
}

@article{Oliveira:2004he,
    author = "Oliveira, J. C. R. E. and Martins, C. J. A. P. and Avelino, P. P.",
    title = "{The Cosmological evolution of domain wall networks}",
    eprint = "hep-ph/0410356",
    archivePrefix = "arXiv",
    doi = "10.1103/PhysRevD.71.083509",
    journal = "Phys. Rev. D",
    volume = "71",
    pages = "083509",
    year = "2005"
}

@article{Hindmarsh:1996xv,
    author = "Hindmarsh, Mark",
    title = "{Analytic scaling solutions for cosmic domain walls}",
    eprint = "hep-ph/9605332",
    archivePrefix = "arXiv",
    reportNumber = "SUSX-TH-96-005",
    doi = "10.1103/PhysRevLett.77.4495",
    journal = "Phys. Rev. Lett.",
    volume = "77",
    pages = "4495--4498",
    year = "1996"
}

@article{Hindmarsh:2002bq,
    author = "Hindmarsh, Mark",
    title = "{Level set method for the evolution of defect and brane networks}",
    eprint = "hep-ph/0207267",
    archivePrefix = "arXiv",
    reportNumber = "SUSX-TH-02-013",
    doi = "10.1103/PhysRevD.68.043510",
    journal = "Phys. Rev. D",
    volume = "68",
    pages = "043510",
    year = "2003"
}

@article{Avelino:2005kn,
    author = "Avelino, P. P. and Martins, C. J. A. P. and Oliveira, J. C. R. E.",
    title = "{One-scale model for domain wall network evolution}",
    eprint = "hep-ph/0507272",
    archivePrefix = "arXiv",
    doi = "10.1103/PhysRevD.72.083506",
    journal = "Phys. Rev. D",
    volume = "72",
    pages = "083506",
    year = "2005"
}

@article{Li:2023gil,
    author = "{Li, Yang and Bian, Ligong and Cai, Rong-Gen and Shu, Jing}",
    title = "{Cosmic Simulations of Axion String-Wall Networks: Probing Dark Matter and Gravitational Waves for Discovery}",
    journal = "{arXiv preprint}",
    note = "{arXiv:2311.02011 [astro-ph.CO]}",
    month = "{11}",
    year = "{2023}"
}

@article{Hiramatsu:2012sc,
    author = "Hiramatsu, Takashi and Kawasaki, Masahiro and Saikawa, Ken'ichi and Sekiguchi, Toyokazu",
    title = "{Axion cosmology with long-lived domain walls}",
    eprint = "1207.3166",
    archivePrefix = "arXiv",
    primaryClass = "hep-ph",
    reportNumber = "ICRR-REPORT-620-2012-9, IPMU12-0140, YITP-12-58",
    doi = "10.1088/1475-7516/2013/01/001",
    journal = "JCAP",
    volume = "01",
    pages = "001",
    year = "2013"
}

@article{ryden1990evolution,
  title={The evolution of networks of domain walls and cosmic strings},
  author={Ryden, Barbara S and Press, William H and Spergel, David N},
  journal={Astrophysical Journal, Part 1 (ISSN 0004-637X), vol. 357, July 10, 1990, p. 293-300.},
  volume={357},
  pages={293--300},
  year={1990}
}

@article{Scherrer:1997sq,
    author = "Scherrer, Robert J. and Vilenkin, Alexander",
    title = "{`Lattice-free' simulations of topological defect formation}",
    eprint = "hep-ph/9709498",
    archivePrefix = "arXiv",
    doi = "10.1103/PhysRevD.58.103501",
    journal = "Phys. Rev. D",
    volume = "58",
    pages = "103501",
    year = "1998"
}

@article{Hiramatsu:2010yz,
    author = "Hiramatsu, Takashi and Kawasaki, Masahiro and Saikawa, Ken'ichi",
    title = "{Gravitational Waves from Collapsing Domain Walls}",
    eprint = "1002.1555",
    archivePrefix = "arXiv",
    primaryClass = "astro-ph.CO",
    reportNumber = "ICRR-REPORT-559-2009-21, IPMU10-0024",
    doi = "10.1088/1475-7516/2010/05/032",
    journal = "JCAP",
    volume = "05",
    pages = "032",
    year = "2010"
}

@article{kawasaki_axion_2015,
	title = {Axion dark matter from topological defects},
	volume = {91},
	issn = {1550-7998, 1550-2368},
	url = {http://arxiv.org/abs/1412.0789},
	doi = {10.1103/PhysRevD.91.065014},
	number = {6},
	urldate = {2024-09-10},
	journal = {Physical Review D},
	author = {Kawasaki, Masahiro and Saikawa, Ken'ichi and Sekiguchi, Toyokazu},
	month = mar,
	year = {2015},
	
}

@article{Kawasaki:2013iha,
    author = "Kawasaki, Masahiro and Yanagida, Tsutomu T. and Yoshino, Kazuyoshi",
    title = "{Domain wall and isocurvature perturbation problems in axion models}",
    eprint = "1305.5338",
    archivePrefix = "arXiv",
    primaryClass = "hep-ph",
    reportNumber = "ICRR-REPORT-653-2013-2, IPMU-13-0104",
    doi = "10.1088/1475-7516/2013/11/030",
    journal = "JCAP",
    volume = "11",
    pages = "030",
    year = "2013"
}

@article{kamionkowski1992planck,
    author = "Kamionkowski, Marc and March-Russell, John",
    title = "{Planck scale physics and the Peccei-Quinn mechanism}",
    eprint = "hep-th/9202003",
    archivePrefix = "arXiv",
    reportNumber = "IASSNS-HEP-92-9, PUPT-92-1309",
    doi = "10.1016/0370-2693(92)90492-M",
    journal = "Phys. Lett. B",
    volume = "282",
    pages = "137--141",
    year = "1992"
}

@article{holman1992solutions,
    author = "Holman, Richard and Hsu, Stephen D. H. and Kephart, Thomas W. and Kolb, Edward W. and Watkins, Richard and Widrow, Lawrence M.",
    title = "{Solutions to the strong CP problem in a world with gravity}",
    eprint = "hep-ph/9203206",
    archivePrefix = "arXiv",
    reportNumber = "NSF-ITP-92-06, CMU-HEP92-05, FERMILAB-PUB-92-034-A, HUTP-92-A011, VAND-TH-92-2",
    doi = "10.1016/0370-2693(92)90491-L",
    journal = "Phys. Lett. B",
    volume = "282",
    pages = "132--136",
    year = "1992"
}

@article{dobrescu1997strong,
    author = "Dobrescu, Bogdan A.",
    title = "{The Strong CP problem versus Planck scale physics}",
    eprint = "hep-ph/9609221",
    archivePrefix = "arXiv",
    reportNumber = "BUHEP-96-30",
    doi = "10.1103/PhysRevD.55.5826",
    journal = "Phys. Rev. D",
    volume = "55",
    pages = "5826--5833",
    year = "1997"
}

@article{barr1992planck,
    author = "Barr, Stephen M. and Seckel, D.",
    title = "{Planck scale corrections to axion models}",
    reportNumber = "BA-92-11",
    doi = "10.1103/PhysRevD.46.539",
    journal = "Phys. Rev. D",
    volume = "46",
    pages = "539--549",
    year = "1992"
}

@inproceedings{dine1992problems,
    author = "Dine, Michael",
    title = "{Problems of naturalness: Some lessons from string theory}",
    booktitle = "{Conference on Topics in Quantum Gravity}",
    eprint = "hep-th/9207045",
    archivePrefix = "arXiv",
    reportNumber = "SCIPP-92-27",
    month = "7",
    year = "1992"
}

@article{garagounis2003scaling,
    author = "Garagounis, Theodore and Hindmarsh, Mark",
    title = "{Scaling in numerical simulations of domain walls}",
    eprint = "hep-ph/0212359",
    archivePrefix = "arXiv",
    reportNumber = "SUSX-TH-02-029",
    doi = "10.1103/PhysRevD.68.103506",
    journal = "Phys. Rev. D",
    volume = "68",
    pages = "103506",
    year = "2003"
}

@article{oliveira2005cosmological,
    author = "Oliveira, J. C. R. E. and Martins, C. J. A. P. and Avelino, P. P.",
    title = "{The Cosmological evolution of domain wall networks}",
    eprint = "hep-ph/0410356",
    archivePrefix = "arXiv",
    doi = "10.1103/PhysRevD.71.083509",
    journal = "Phys. Rev. D",
    volume = "71",
    pages = "083509",
    year = "2005"
}

@article{avelino2005one,
    author = "Avelino, P. P. and Martins, C. J. A. P. and Oliveira, J. C. R. E.",
    title = "{One-scale model for domain wall network evolution}",
    eprint = "hep-ph/0507272",
    archivePrefix = "arXiv",
    doi = "10.1103/PhysRevD.72.083506",
    journal = "Phys. Rev. D",
    volume = "72",
    pages = "083506",
    year = "2005"
}

@article{leite2011scaling,
    author = "Oliveira, M. F. and Martins, C. J. A. P.",
    title = "{Scaling properties of multitension domain wall networks}",
    eprint = "1503.00234",
    archivePrefix = "arXiv",
    primaryClass = "hep-ph",
    doi = "10.1103/PhysRevD.91.043527",
    journal = "Phys. Rev. D",
    volume = "91",
    number = "4",
    pages = "043527",
    year = "2015"
}

@article{leite2013accurate,
    author = "Leite, A. M. M. and Martins, C. J. A. P. and Shellard, E. P. S.",
    title = "{Accurate Calibration of the Velocity-dependent One-scale Model for Domain Walls}",
    eprint = "1206.6043",
    archivePrefix = "arXiv",
    primaryClass = "hep-ph",
    doi = "10.1016/j.physletb.2012.11.070",
    journal = "Phys. Lett. B",
    volume = "718",
    pages = "740--744",
    year = "2013"
}

@article{martins2016extending,
    author = "Martins, C. J. A. P. and Rybak, I. Yu. and Avgoustidis, A. and Shellard, E. P. S.",
    title = "{Extending the velocity-dependent one-scale model for domain walls}",
    eprint = "1602.01322",
    archivePrefix = "arXiv",
    primaryClass = "hep-ph",
    doi = "10.1103/PhysRevD.93.043534",
    journal = "Phys. Rev. D",
    volume = "93",
    number = "4",
    pages = "043534",
    year = "2016"
}

@article{hiramatsu2013axion,
    author = "Hiramatsu, Takashi and Kawasaki, Masahiro and Saikawa, Ken'ichi and Sekiguchi, Toyokazu",
    title = "{Axion cosmology with long-lived domain walls}",
    eprint = "1207.3166",
    archivePrefix = "arXiv",
    primaryClass = "hep-ph",
    reportNumber = "ICRR-REPORT-620-2012-9, IPMU12-0140, YITP-12-58",
    doi = "10.1088/1475-7516/2013/01/001",
    journal = "JCAP",
    volume = "01",
    pages = "001",
    year = "2013"
}

@article{hiramatsu2014estimation,
    author = "Hiramatsu, Takashi and Kawasaki, Masahiro and Saikawa, Ken'ichi",
    title = "{On the estimation of gravitational wave spectrum from cosmic domain walls}",
    eprint = "1309.5001",
    archivePrefix = "arXiv",
    primaryClass = "astro-ph.CO",
    reportNumber = "ICRR-REPORT-659-2013-8, IPMU13-0182, YITP-13-87",
    doi = "10.1088/1475-7516/2014/02/031",
    journal = "JCAP",
    volume = "02",
    pages = "031",
    year = "2014"
}

@article{kawasaki2011study,
    author = "Kawasaki, Masahiro and Saikawa, Ken'ichi",
    title = "{Study of gravitational radiation from cosmic domain walls}",
    eprint = "1102.5628",
    archivePrefix = "arXiv",
    primaryClass = "astro-ph.CO",
    reportNumber = "ICRR-REPORT-581-2010-14, IPMU11-0032",
    doi = "10.1088/1475-7516/2011/09/008",
    journal = "JCAP",
    volume = "09",
    pages = "008",
    year = "2011"
}

@article{figueroa2021art,
    author = "Figueroa, Daniel G. and Florio, Adrien and Torrenti, Francisco and Valkenburg, Wessel",
    title = "{The art of simulating the early Universe -- Part I}",
    eprint = "2006.15122",
    archivePrefix = "arXiv",
    primaryClass = "astro-ph.CO",
    doi = "10.1088/1475-7516/2021/04/035",
    journal = "JCAP",
    volume = "04",
    pages = "035",
    year = "2021"
}

@article{figueroa2023cosmolattice,
    author = "Figueroa, Daniel G. and Florio, Adrien and Torrenti, Francisco and Valkenburg, Wessel",
    title = "{CosmoLattice: A modern code for lattice simulations of scalar and gauge field dynamics in an expanding universe}",
    eprint = "2102.01031",
    archivePrefix = "arXiv",
    primaryClass = "astro-ph.CO",
    doi = "10.1016/j.cpc.2022.108586",
    journal = "Comput. Phys. Commun.",
    volume = "283",
    pages = "108586",
    year = "2023"
}

@article{king2017unified,
    author = "King, S. F.",
    title = "{Unified Models of Neutrinos, Flavour and CP Violation}",
    eprint = "1701.04413",
    archivePrefix = "arXiv",
    primaryClass = "hep-ph",
    doi = "10.1016/j.ppnp.2017.01.003",
    journal = "Prog. Part. Nucl. Phys.",
    volume = "94",
    pages = "217--256",
    year = "2017"
}

@article{xing2020flavor,
    author = "Xing, Zhi-zhong",
    title = "{Flavor structures of charged fermions and massive neutrinos}",
    eprint = "1909.09610",
    archivePrefix = "arXiv",
    primaryClass = "hep-ph",
    doi = "10.1016/j.physrep.2020.02.001",
    journal = "Phys. Rept.",
    volume = "854",
    pages = "1--147",
    year = "2020"
}

@article{ibanez1992discrete,
    author = "Ibanez, Luis E. and Ross, Graham G.",
    title = "{Discrete gauge symmetries and the origin of baryon and lepton number conservation in supersymmetric versions of the standard model}",
    reportNumber = "CERN-TH-6111-91",
    doi = "10.1016/0550-3213(92)90195-H",
    journal = "Nucl. Phys. B",
    volume = "368",
    pages = "3--37",
    year = "1992"
}

@article{larsson1997evading,
    author = "Larsson, Sebastian E. and Sarkar, Subir and White, Peter L.",
    title = "{Evading the cosmological domain wall problem}",
    eprint = "hep-ph/9608319",
    archivePrefix = "arXiv",
    reportNumber = "OUTP-96-11-P",
    doi = "10.1103/PhysRevD.55.5129",
    journal = "Phys. Rev. D",
    volume = "55",
    pages = "5129--5135",
    year = "1997"
}

@article{gelmini1989cosmology,
    author = "Gelmini, Graciela B. and Gleiser, Marcelo and Kolb, Edward W.",
    title = "{Cosmology of Biased Discrete Symmetry Breaking}",
    reportNumber = "NSF-ITP-88-148, FERMILAB-PUB-88-151-A",
    doi = "10.1103/PhysRevD.39.1558",
    journal = "Phys. Rev. D",
    volume = "39",
    pages = "1558",
    year = "1989"
}

@article{vilenkin1981gravitational,
    author = "Vilenkin, A.",
    title = "{Gravitational Field of Vacuum Domain Walls and Strings}",
    doi = "10.1103/PhysRevD.23.852",
    journal = "Phys. Rev. D",
    volume = "23",
    pages = "852--857",
    year = "1981"
}

@article{Bian:2022qbh,
    author = "Bian, Ligong and Ge, Shuailiang and Li, Changhong and Shu, Jing and Zong, Junchao",
    title = "{Domain wall network: A dual solution for gravitational waves and Hubble tension?}",
    eprint = "2212.07871",
    archivePrefix = "arXiv",
    primaryClass = "hep-ph",
    doi = "10.1007/s11433-024-2436-4",
    journal = "Sci. China Phys. Mech. Astron.",
    volume = "67",
    number = "11",
    pages = "110413",
    year = "2024"
}

@article{dankovsky2024revisiting,
    author = "Dankovsky, I. and Babichev, E. and Gorbunov, D. and Ramazanov, S. and Vikman, A.",
    title = "{Revisiting evolution of domain walls and their gravitational radiation with CosmoLattice}",
    eprint = "2406.17053",
    archivePrefix = "arXiv",
    primaryClass = "astro-ph.CO",
    doi = "10.1088/1475-7516/2024/09/047",
    journal = "JCAP",
    volume = "09",
    pages = "047",
    year = "2024"
}

@article{Li:2023yzq, author = "Li, Yang and Jia, Yongtao and Bian, Ligong", title = "{Numerical simulation of domain wall and first-order phase transition in an expanding universe}", eprint = "2304.05220", archivePrefix = "arXiv", primaryClass = "hep-ph", doi = "10.1088/1475-7516/2025/02/038", journal = "JCAP", volume = "02", pages = "038", year = "2025" }

@article{Gelmini:2020bqg,
    author = "Gelmini, Graciela B. and Pascoli, Silvia and Vitagliano, Edoardo and Zhou, Ye-Ling",
    title = "{Gravitational wave signatures from discrete flavor symmetries}",
    eprint = "2009.01903",
    archivePrefix = "arXiv",
    primaryClass = "hep-ph",
    doi = "10.1088/1475-7516/2021/02/032",
    journal = "JCAP",
    volume = "02",
    pages = "032",
    year = "2021"
}

@article{Gelmini:2021yzu,
    author = "Gelmini, Graciela B. and Simpson, Anna and Vitagliano, Edoardo",
    title = "{Gravitational waves from axionlike particle cosmic string-wall networks}",
    eprint = "2103.07625",
    archivePrefix = "arXiv",
    primaryClass = "hep-ph",
    doi = "10.1103/PhysRevD.104.L061301",
    journal = "Phys. Rev. D",
    volume = "104",
    number = "6",
    pages = "061301",
    year = "2021"
}

@article{Gelmini:2022nim,
    author = "Gelmini, Graciela B. and Simpson, Anna and Vitagliano, Edoardo",
    title = "{Catastrogenesis: DM, GWs, and PBHs from ALP string-wall networks}",
    eprint = "2207.07126",
    archivePrefix = "arXiv",
    primaryClass = "hep-ph",
    doi = "10.1088/1475-7516/2023/02/031",
    journal = "JCAP",
    volume = "02",
    pages = "031",
    year = "2023"
}

@article{Gelmini:2023ngs,
    author = "Gelmini, Graciela B. and Hyman, Jonah and Simpson, Anna and Vitagliano, Edoardo",
    title = "{Primordial black hole dark matter from catastrogenesis with unstable pseudo-Goldstone bosons}",
    eprint = "2303.14107",
    archivePrefix = "arXiv",
    primaryClass = "hep-ph",
    doi = "10.1088/1475-7516/2023/06/055",
    journal = "JCAP",
    volume = "06",
    pages = "055",
    year = "2023"
}

@article{Larsson:1996sp,
    author = "Larsson, Sebastian E. and Sarkar, Subir and White, Peter L.",
    title = "{Evading the cosmological domain wall problem}",
    eprint = "hep-ph/9608319",
    archivePrefix = "arXiv",
    reportNumber = "OUTP-96-11-P",
    doi = "10.1103/PhysRevD.55.5129",
    journal = "Phys. Rev. D",
    volume = "55",
    pages = "5129--5135",
    year = "1997"
}

\end{document}